\documentclass[12pt,preprint]{aastex}
\usepackage{amsmath}
\usepackage{graphicx}
\usepackage{enumitem}
\usepackage{color}

\shorttitle{Resonant Absorption in a Solar Prominence - Simulations}
\shortauthors{P. Antolin et al.}

\begin{document}

\title{Resonant Absorption of Transverse Oscillations and Associated Heating in a Solar Prominence. II- Numerical aspects}

\author{P. Antolin\altaffilmark{1}, T. J. Okamoto\altaffilmark{2,8}, B. De Pontieu\altaffilmark{3,4}, H. Uitenbroek\altaffilmark{5}, T. Van Doorsselaere\altaffilmark{6}, T. Yokoyama\altaffilmark{7}}
\affil{\altaffilmark{1}National Astronomical Observatory of Japan, Osawa, Mitaka, Tokyo 181-8588, Japan\\
\altaffilmark{2}ISAS/JAXA, Sagamihara, Kanagawa 252-5210, Japan\\
\altaffilmark{3}Lockheed Martin Solar and Astrophysics Laboratory, 3251 Hanover Street, Palo Alto, CA 94304, USA\\
\altaffilmark{4}Institute of Theoretical Astrophysics, University of Oslo, P.O. Box 1029 Blindern, N-0315 Oslo, Norway\\
\altaffilmark{5}National Solar Observatory, PO Box 62, sunspot, NM 88349, USA\\
\altaffilmark{6}Centre for Mathematical Plasma Astrophysics, Mathematics Department, KU Leuven, Celestijnenlaan 200B bus 2400, B-3001 Leuven, Belgium\\
\altaffilmark{7}The University of Tokyo, Hongo, Bunkyo-ku, Tokyo 113-0033, Japan\\
\altaffilmark{8}Current address: STEL, Nagoya University, Aichi 464-8601, Japan}
\email{patrick.antolin@nao.ac.jp}

\begin{abstract}

Transverse magnetohydrodynamic (MHD) waves are ubiquitous in the solar atmosphere and may be responsible for generating the Sun's million-degree outer atmosphere. However, direct evidence of the dissipation process and heating from these waves remains elusive. Through advanced numerical simulations combined with appropriate forward modeling of a prominence flux tube, we provide the observational signatures of transverse MHD waves in prominence plasmas. We show that these signatures are characterized by thread-like substructure, strong transverse dynamical coherence, an out-of-phase difference between plane-of-the-sky motions and LOS velocities, and enhanced line broadening and heating around most of the flux tube. A complex combination between resonant absorption and Kelvin-Helmholtz instabilities (KHI) takes place in which the KHI extracts the energy from the resonant layer and dissipates it through vortices and current sheets, which rapidly degenerate into turbulence. An inward enlargement of the boundary is produced in which the turbulent flows conserve the characteristic dynamics from the resonance, therefore guaranteeing detectability of the resonance imprints. We show that the features described in the accompanying paper \citep{Okamoto_etal_2015} through coordinated \emph{Hinode} and \emph{IRIS} observations match well the numerical results.

\end{abstract}


\section{Introduction}

Observational reports of periodic transverse displacements of chromospheric and coronal structures have increased in the last decade, especially due to the advance in instrumentation allowing higher sensitivity and spatial, temporal and spectral resolution. Initially reported through EUV imaging as large disturbances of coronal loops in the aftermath of flares with \textit{TRACE} \citep{Nakariakov_1999Sci...285..862N}, they are now observed more frequently with \textit{SDO} \citep{McIntosh_2011Natur.475..477M,Threlfall_2013AA...556A.124T,Thurgood_2014ApJ...790L...2T}. Observations with the Solar Optical Telescope \citep[SOT,][]{Tsuneta_2008SoPh..249..167T} on board \textit{Hinode} \citep{Kosugi_2007SoPh..243....3K}, together with other ground-based facilities providing high spatial resolution imaging, have established that such low amplitude periodic disturbances are ubiquitous in the chromosphere  \citep{DePontieu_2007Sci...318.1574D,DePontieu_etal_2007PASJ...59S.655D,Kuridze_2012ApJ...750...51K,Morton_1012NatCommun...3...1315} and in the corona, through observations of prominences or coronal rain  \citep{Okamoto_2007Sci...318.1577O,Terradas_2008ApJ...678L.153T,Lin_2009ApJ...704..870L,Lin_2011SSRv..158..237L,Antolin_Verwichte_2011ApJ...736..121A,Arregui_2012LRSP....9....2A,Hillier_2013ApJ...779L..16H,Schmieder_2013ApJ...777..108S}. Periodic spectral disturbances in Doppler velocities have also been reported, frequently associated with corresponding swaying in intensity of chromospheric or coronal structures \citep{Tomczyk_2007Sci...317.1192T,Erdelyi_2008AA...489L..49E,VanDoorsselaere_etal_2008AA...487L..17V,Tomczyk_2009ApJ...697.1384T,Lin_2009ApJ...704..870L,Tian_2012ApJ...759..144T,Schmieder_2013ApJ...777..108S}. The ubiquity of such motions and also of torsional components has recently been further established in spicules, by the \textit{Interface Region Imaging Spectrograph} \citep[\textit{IRIS,}][]{DePontieu_2014SoPh..289.2733D,DePontieu_2014Sci...346D.315D,Rouppe_2015ApJ...799L...3R,Skogsrud_2014ApJ...795L..23S}. 

Such periodic transverse disturbances usually have low amplitudes (a few km s$^{-1}$), periods in the order of a few to tens of minutes and usually exhibit damping signatures. General consensus exists on the interpretation of such atmospheric disturbances as Alfv\'enic waves, specifically transverse MHD waves \citep[e.g.,][and references therein]{Oliver_2009SSRv..149..175O,Mackay_2010SSRv..151..333M,Arregui_2012LRSP....9....2A,Mathioudakis_2013SSRv..175....1M}, also known as kink waves or kink surface Alfv\'en waves \citep{Goossens_2002AA...394L..39G,Goossens_2012ApJ...753..111G}. The importance of these Alfv\'enic waves is set partly on the close connection between their characteristics and the medium in which they propagate, thereby serving as seismological probes of the magnetic field \citep{Nakariakov_Ofman_2001AA...372L..53N}, the loop radial structure \citep{Aschwanden_Nightingale_2003ApJ...598.1375A}, the vertical density scale height  \citep{Andries_2005ApJ...624L..57A} and Alfv\'en transit times \citep{Arregui_2007AA...466.1145A}. Alfv\'enic waves are further important due to their ability to carry significant amounts of energy over long distances, thereby potentially playing an important role in coronal heating and solar wind acceleration \citep{Alfven_1947MNRAS.107..211A,Uchida_1974SoPh...35..451U,Antolin_2010ApJ...712..494A,vanBallegooijen_etal_2011ApJ...736....3V,Matsumoto_Suzuki_2014MNRAS.440..971M}. These waves are often observed to damp on short spatial and temporal scales  \citep{Aschwanden_1999ApJ...520..880A,Pascoe_2010ApJ...711..990P,Pascoe_etal_2011ApJ...731...73P}. However, it is still unclear whether the damping usually leads to significant dissipation, and thus heating \citep[see][for a contemporary review]{Parnell_DeMoortel_2012RSPTA.370.3217P}.

The dissipation process of transverse MHD waves in an inhomogeneous medium has been a subject of active research over several decades \citep[e.g.,][]{Ionson_1978ApJ...226..650I,Hollweg_1990ApJ...349..335H,Sakurai_1991SoPh..133..227S,Goossens_1992SoPh..138..233G}. Theoretically it is expected that the transverse, bulk motion is subject to damping by resonant absorption in an inhomogeneous flux tube, such as a prominence thread \citep{Arregui_2008ApJ...682L.141A,Arregui_2011SSRv..158..169A,Soler_2012AA...546A..82S}. An analogous mechanism known as field line resonance exists in the magnetosphere \citep{Southwood_Hughes_1983SSRv...35..301S}. Resonant absorption occurs at the tube's boundary layer, where the phase speed of the trapped (anisotropic) kink wave matches that of the external (isotropic) Alfv\'en waves. The kink wave (with azimuthal wave number $m=1$), whose initial energy is mostly transverse, mode converts into azimuthal waves locally resembling torsional Alfv\'en waves, which are finely localized around the tube's boundary layer \citep{Verth_2010ApJ...718L.102V,Arregui_2011AA...533A..60A}. Azimuthal motions are thus amplified and introduce velocity shear, making them prone to be unstable to the Kelvin-Helmholtz instability (KHI) \citep{Heyvaerts_1983AA...117..220H,Soler_2010ApJ...712..875S}. The KHI deforms the boundary layer and leads to enhanced dissipation of the wave energy into heat in thin, turbulent current sheets \citep{Uchimoto_1991SoPh..134..111U,Karpen_1993ApJ...403..769K,Poedts_1997SoPh..172...45P, Ziegler_1997AA...327..854Z,Ofman_1998SoPh..183...97O,Terradas_2008ApJ...687L.115T}. Apart from chromospheric and coronal heating, the importance of this instability has been stressed in magnetic reconnection studies \citep{Lapenta_2003SoPh..214..107L} and in the mixing of plasma between two magnetically confined plasma environments, such as closed and open magnetic field lines \citep[relevant for solar wind studies][]{Andries_2001AA...368.1083A} or across the magnetospheric boundary  \citep{Fujimoto_1994JGR....99.8601F,Fujimoto_2006SSRv..122....3F}.

Recent simulations by \citet{Antolin_2014ApJ...787L..22A} show that even small amplitude oscillations can lead to the KHI, whose vortices combined with line-of-sight (LOS) effects result in strand-like structures within coronal loops, but at scales that are difficult to resolve with current coronal instruments. This study opens the question of whether an important part of the observed strand-like structure in loops \citep{Ofman_Wang_2008AA...482L...9O,Reale_2010LRSP....7....5R,Antolin_Verwichte_2011ApJ...736..121A,Antolin_Rouppe_2012ApJ...745..152A,Brooks_2013ApJ...772L..19B,Peter_2013AA...556A.104P,Scullion_2014ApJ...797...36S} or thread-like structure in prominences \citep{Lin_2011SSRv..158..237L,Okamoto_2007Sci...318.1577O,Terradas_2008ApJ...678L.153T,Vial_Engvold_2015ASSL..415.....V} could be linked to transverse MHD waves through their ubiquity and the generation of the KHI. 

In this series of two papers, unique coordinated observations with \emph{Hinode}/SOT of \emph{Hinode} and \emph{IRIS} are used to provide evidence for a telltale sign of resonant absorption and associated KHI and heating in an active region solar prominence \citep[][hereafter Paper~1]{Okamoto_etal_2015}. In the present paper details of the three-dimensional MHD simulations and forward modeling first introduced in Paper~1 are given, and we further provide a complete description of the observational signatures expected for transverse MHD waves. As discussed in this paper most of the obtained signatures in the numerical model are also expected in other solar atmospheric structures such as coronal loops \citep{Antolin_2014ApJ...787L..22A} and spicules, and may be considered general to a certain extent. 

The paper is organized as follows. In section~\ref{model} the details of the numerical model are provided. The results of the forward modeling of the numerical results are given in section~\ref{forward}. Results are discussed in section~\ref{discussion} and conclusions in section~\ref{conclusions}.

\section{Numerical Model}\label{model}

\begin{figure}[!ht]
\begin{center}
$\begin{array}{c}
\includegraphics[scale=1]{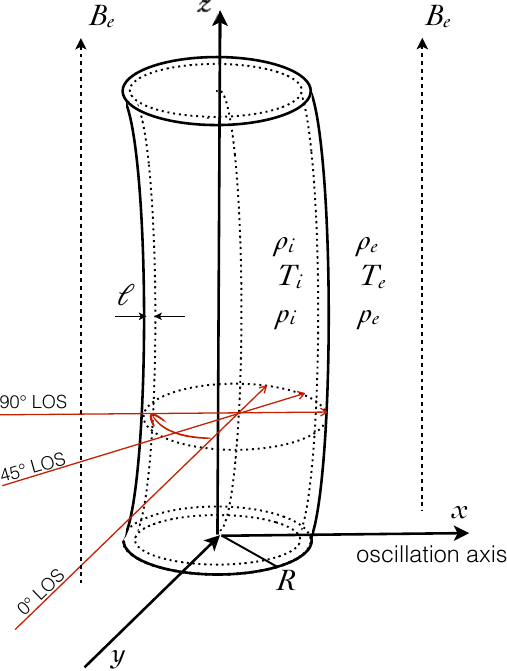}
\end{array}$
\caption{Sketch of the prominence flux tube subject to a transverse MHD wave in our numerical model. The axis of the flux tube is oriented along the $z$ axis. The transverse perturbation (and the subsequent oscillation) is directed along the $x$-axis. The LOS for the forward modeling is in the $(x,y)$ plane. The $0^{\circ}$ and $90^{\circ}$ LOS rays are directed along the positive $y-axis$ and $x$-axis, respectively, as shown in the figure. Variables $\rho$, $T$, $p$ and $B$ denote, respectively, the density, temperature, pressure and magnetic field within (subindex $i$) or outside (subindex $e$) the flux tube. $R$ and $l$ denote, respectively, the radius of the flux tube and the width of the inhomogeneous boundary layer.
\label{fig_init}}
\end{center}
\end{figure}

To interpret the observations in Paper~1 we performed 3D MHD simulations of a prominence flux tube oscillating with a fundamental kink mode. The numerical model is similar to that of the previous work  \cite{Antolin_2014ApJ...787L..22A} for the coronal case, and we adapt it to the prominence case. We considered a flux tube with physical quantities usually found in solar prominences \citep{Tandberg-Hansen_1995ASSL..199.....T,Vial_Engvold_2015ASSL..415.....V}. We take a number density and temperature contrast of 10 and 1/100, respectively, with respect to the external values, which we set to $10^{9}$~cm$^{-3}$ ($10^{15}$~m$^{-3}$) and $10^{6}$~K. The external plasma is assumed to be fully ionized with a beta value of 0.01, leading to an external magnetic field value of 18.6~G (1.86~mT) and a slightly larger value within the flux tube in order to keep a total pressure balance. The external and internal thermodynamic values are linked smoothly as in \cite{Antolin_2014ApJ...787L..22A}, defining a specific width $l$ for the boundary layer (see Fig.~\ref{fig_init}). Here we adopt a width of $l/R\approx0.4$, where $R$ denotes the radius of the tube. The dimensions of the flux tube are to some extent constrained by the observations. AIA reveals that the length of the prominence is larger than $150,000$~km and exists over a height range of roughly $30,000$~km. However, a strong LOS projection effect is expected. Also, dynamic coherence is only detected over a few 1,000 km width in the vertical direction (fig. 2 in Paper 1), suggesting that the prominence is likely to be composed of several dynamically independent flux tubes. We thus choose to take the length of the tube as $200~R$ and we set $R=1,000$~km. 

At time $t=0$ the flux tube is given a perturbation in velocity mimicking the fundamental kink mode (with longitudinal wavenumber $k\approx0.015/R$) as described in \cite{Antolin_2014ApJ...787L..22A}. The amplitude of the pulse is $v_{0}\approx 8$~ km s$^{-1}$, a value in the low end of those being detected with SOT and IRIS. The kink phase speed is $c_{k} \approx 776$~km s$^{-1}$, and therefore the perturbation is largely linear.

The numerical simulations were performed with the CIP-MOCCT scheme \citep{Kudoh_1999_CFD.8}, which solves the 3D MHD equations including resistivity and viscosity. Several simulations were performed varying the values of resistivity and viscosity, and the number of grid points. The base run consists of a numerical box with $512\times256\times50$ points in the $x, y$ and $z$ directions, respectively, where $x$ and $y$ denote the parallel and perpendicular directions of oscillation, respectively, and $z$ is the longitudinal direction. Higher resolution runs with double number of grid points in every direction were also performed. The values of resistivity and viscosity are set to small constant values everywhere, ensuring large enough Lundquist (and viscous) Reynolds numbers of $10^{4}-10^{5}$ in the base run, and around $10^{6}-10^{7}$ in the large resolution run. While a considerable increase of fine scale structure is present in the high resolution runs, the main results are left unchanged. This is expected since the effects from resonant absorption stabilise for high Reynolds numbers \citep{Kappraff_Tataronis_1977JPlPh..18..209K,Poedts_Kerner_1991PhRvL..66.2871P,Ofman_Davila_1995JGR...10023427O,Tirry_Goossens_1996ApJ...471..501T,VanDoorsselaere_2004ApJ...606.1223V, Terradas_etal_2006ApJ...642..533T}. The results presented here are from the base run unless explicitly stated.

Thanks to the symmetric properties of the kink mode only half the plane in $y$ and $z$ is modeled, and we set symmetric boundary conditions in all boundary planes except for the $x$ boundary planes, where periodic boundary conditions are imposed. In order to minimize the influence from side boundary conditions the spatial grids in $x$ and $y$ are non-uniform, with exponentially increasing values for distances beyond the maximum displacement. The maximum distance in $x$ and $y$ from the center is $\approx 16~R$. The spatial resolution at the loop's location is $0.0156~R$, and a tenth of this value for the large resolution run.

\begin{figure*}[!ht]
\begin{center}
$\begin{array}{cc}
\includegraphics[scale=0.5]{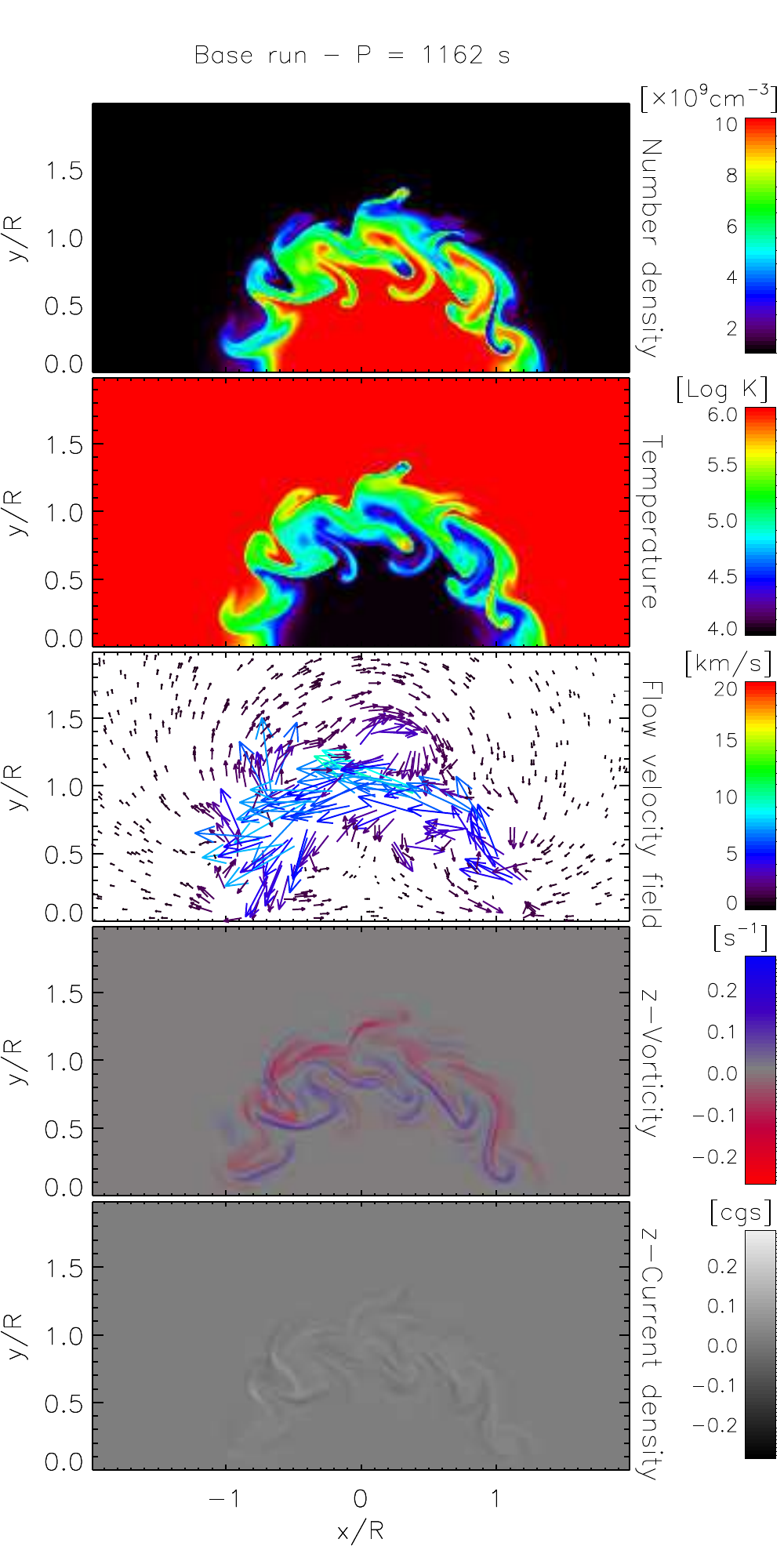} &
\includegraphics[scale=0.5]{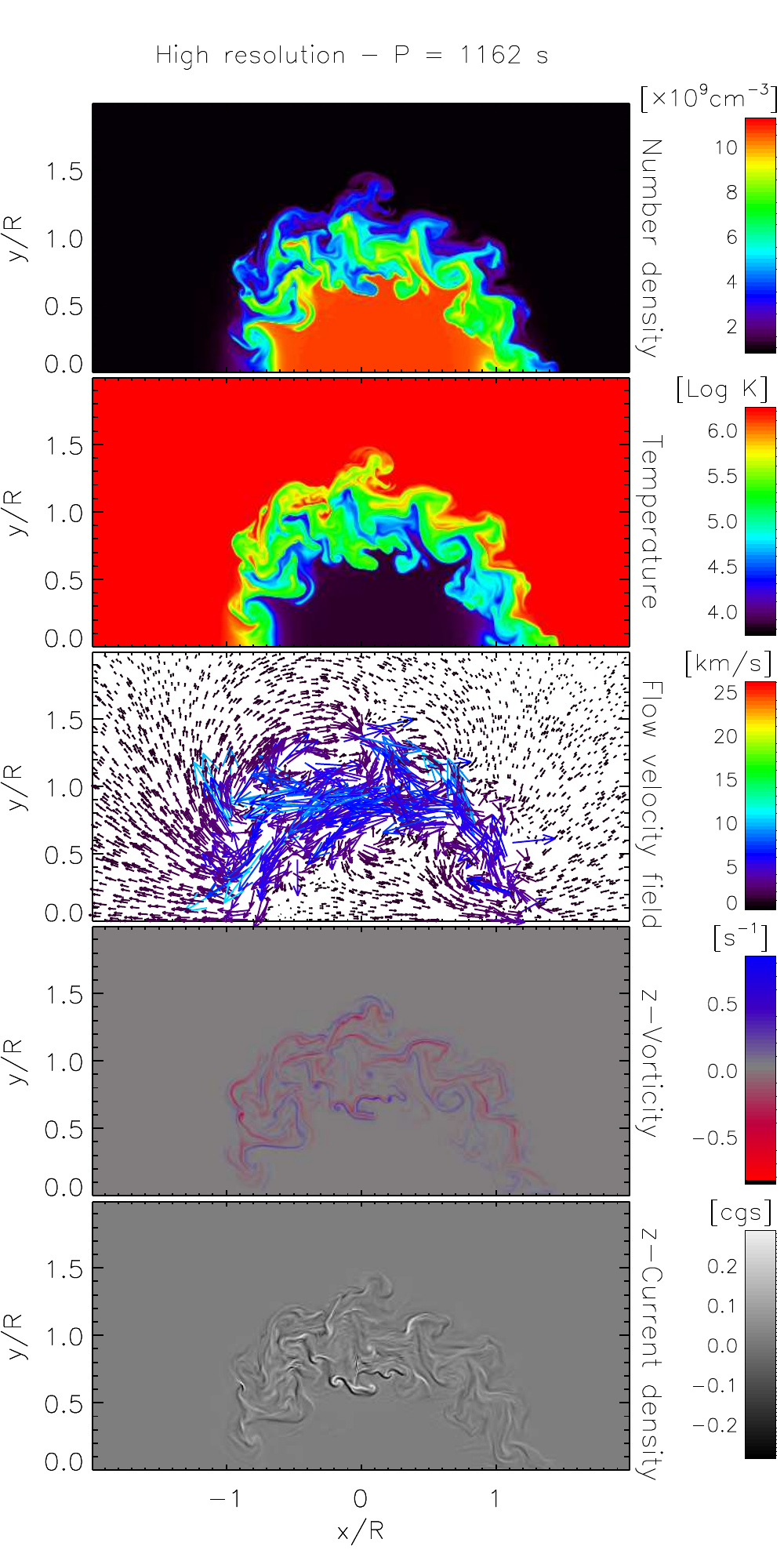}
\end{array}$
\caption{Snapshot of the cross-section of several physical quantities at the apex of the prominence thread at a time in the simulation of $t=1162~$s, corresponding to roughly 2 oscillation periods. From top to bottom the panels in both columns correspond to number density, temperature (logarithm), flow velocity field, and the longitudinal ($z$-) components of the vorticity and current density. The left and right panels correspond, respectively, to the base and high resolution runs. An animation of this figure for the high resolution run is available online (movie~1). See also movie~B in Paper~1 for the corresponding animation of the base run.
\label{fig_model}}
\end{center}
\end{figure*}

\subsection{Resonant absorption in the numerical simulation}\label{RA}
Following the initial kink the tube oscillates with a period of 525~s, close to the fundamental period of $2L/c_{k}$. The maximum displacement of the tube is 710~km and the damping time of the oscillation is roughly 840~s, a value 16\% smaller than that predicted by resonant absorption theory in the linear regime. As the tube is displaced transversely by the kink mode resonant absorption sets in immediately, as expected from theory \citep{Poedts_1990CoPhC..59...95P}. This can be seen in movie~1, as black and white rings in the $z$-vorticity panel, occurring around the boundary layer. In the boundary layer, where the Alfv\'en speed is close to the kink speed, the nature of the kink quasi-mode is modified and becomes similar to that of a torsional Alfv\'en wave. Energy is transferred from transverse motions to azimuthal motions around the resonant layer, seen as bright or dark rings in the $z$-vorticity panel. The process of resonant absorption has been extensively studied in the past and has been shown to be an efficient energy conversion mechanism between waves \citep{Goossens_2002AA...394L..39G,Goossens_2006RSPTA.364..433G,VanDoorsselaere_2004ApJ...606.1223V,Goossens_2011SSRv..158..289G,Pascoe_2010ApJ...711..990P}. It has been further shown to be a robust mechanism, with little dependence on the degree of density inhomogeneity within the tube \citep{DeMoortel_Nakariakov_2012RSPTA.370.3193D,Pascoe_etal_2011ApJ...731...73P,Terradas_2008ApJ...679.1611T}.

\subsection{Kelvin-Helmholtz feedback from resonant absorption}\label{KHI}

Apart from resonant absorption, the KHI sets in rapidly, just one period after the initial kink, as seen in movie~1. The instability is generated by the shear flows at the edges of the loop, produced by the kink mode (from the dipole-like azimuthal flow outside the tube and the transverse displacement of the tube). Such flows have been shown to generate the instability in photospheric and coronal conditions \citep{Karpen_1993ApJ...403..769K,Ofman_1994GeoRL..21.2259O,Poedts_1997SoPh..172...45P, Ziegler_1997AA...327..854Z,Terradas_2008ApJ...687L.115T}, for small amplitude kink modes \citep{Antolin_2014ApJ...787L..22A} matching the currently observed transverse motion in the solar atmosphere \citep{Okamoto_2007Sci...318.1577O,Tomczyk_2007Sci...317.1192T,Erdelyi_2008AA...489L..49E,VanDoorsselaere_etal_2008AA...487L..17V,Terradas_2008ApJ...678L.153T,Lin_2009ApJ...704..870L,Lin_2011SSRv..158..237L,McIntosh_2011Natur.475..477M,Tian_2012ApJ...759..144T,Antolin_Verwichte_2011ApJ...736..121A,Arregui_2012LRSP....9....2A,Hillier_2013ApJ...779L..16H,Schmieder_2013ApJ...777..108S}. Resonant absorption helps the onset of the instability by increasing the amplitude of the azimuthal flow in the boundary layer and therefore that of the shear flow. Here we show that the instability can easily be obtained in prominences. The fast growth of the unstable modes is guaranteed by the large density contrast and the small longitudinal wavenumber, as shown in  \citet{Antolin_2014ApJ...787L..22A}. In the present case, application of formula 2 in the previous paper predicts the most unstable mode with azimuthal wavenumber $m = 3$. In the base and high resolution run 3 and 5 large eddies, respectively, are initially obtained, close to the theoretical results from linear analysis. 

The instability produces vortices (eddies in 2D cross-sections) that develop in the plane perpendicular to the axis of the flux tube, as shown in Fig.~\ref{fig_model}. Such coherent structure rapidly degenerates in turbulence as small scales are produced from the continuous shear flows. The vortices' lifetime is about 1 period, and they are continuously generated and destroyed during the entire simulation. It is important to notice that the vortices do not only affect the boundary of the flux tube but also penetrate deeply towards the core. This can be seen clearly in movie~1 and Fig.~\ref{fig_model}. As a result, the ensuing turbulence enlarges the boundary layer and rapidly drifts inwards towards the center of the tube. Strong deformation of the tube's cross section is clearly visible within one period from the onset of the KHI: the initially dense cool core is significantly reduced to roughly half its original size, and the boundary layer has enlarged to roughly the same size of the core ($l/R\approx1$). The enlarged boundary layer has a temperature around $10^{5}$ K, meaning that most of the tube has been subject to heating. The nature of this heating is not only viscous but also comes from ohmic dissipation. Indeed, we obtain current sheets at the vortices \citep{Ofman_2009ApJ...694..502O,Antolin_2014ApJ...787L..22A}, therefore spanning along most of the prominence flux tube, due to the relatively strong shear in the magnetic field. 

Component magnetic reconnection can occur at such current sheets, as has previously been shown \citep{Lapenta_2003SoPh..214..107L}. The heating that ensues from viscous and ohmic dissipation can be strong and impulsive at the start of the KHI exhibiting various short lived bursts, but rapidly sets in to a constant and continuous tendency. The KHI effectively extracts the kinetic and magnetic energy in the boundary layer deposited by the resonant absorption mechanism, and converts it into heat through viscous and ohmic dissipation from the turbulence and current sheet generation. The regions around the resonant layer affected by the KHI heat up rapidly. This means that most of the tube is heated from chromospheric to transition region temperatures in the time scale of one period. In the high resolution run the deformation and inward enlargement of the boundary layer is more severe. This is because of the enhanced turbulence due to the significant increase of fine-scale structure. The number of vortices and current sheets increases leading to wider and faster heating from a more efficient energy cascade towards smaller scales. The timescale of the heating of most of the flux tube to transition region temperatures is decreased by a few minutes compared to the base run. Only the inner core is more slowly and gradually heated. It is important to notice that the energy transfer from the resonant layer towards the inner parts of the flux tube in our model does not depend on the perpendicular thermal conduction, which is expected to be very small even in the case of prominences. This transfer is produced by the constant generation of small-scale current sheets and vortices in the turbulent flows that occur in the perpendicular magnetic field direction (but which never cross the magnetic field).

\subsection{Resistivity in the numerical model}\label{resist}

The Lundquist number in the base run is on the order of $10^{4} - 10^{5}$, and $10^{6} - 10^{7}$ in the high resolution run. On the other hand, the resistivity in the solar corona is expected to be many orders of magnitude less than the numerical resistivity, indicating much higher Lundquist numbers in the solar corona than those considered here. However, the development of the KHI and phase mixing leads to the rapid formation of gradients on smaller and smaller scales thereby lessening the impact of the resistivity issue. In addition the prominence consists of chromospheric material in which neutral particles naturally occur. As a result, the resistivity in chromospheric plasmas can be on the order of that in numerical models, especially in the presence of ambipolar diffusion \citep{MartinezSykora_2012ApJ...753..161M}.

\subsection{Limitations of the numerical model}

In this numerical model gravity is not included, implying no density stratification along the flux tube. Thermal conduction and radiation are not included either. On the geometrical side, the flux tube is straight, hence effects from curvature, expansion or coherent twist are also neglected. Finally, longitudinal flows are not taken into account either. While each of these factors is present in the observed prominence to a certain degree, we believe that they do not contribute substantially to the physical mechanisms discussed here. Each of these mechanisms has first order contributions in the longitudinal direction and are only second order in the transverse direction, which is where the important dynamics for the present investigation take place. Furthermore the timescales of resonant absorption or the KHI are significantly faster than those of thermal conduction or radiation. Prominences such as the one observed here are mostly horizontal with respect to the solar surface, implying a minor role for stratification, expansion or curvature. Flows can play an important role on damping or enhancement of transverse oscillations \citep{Gruszecki_2008AA...488..757G,Terradas_2010AA...515A..46T}, which can limit the efficiency of resonant absorption and the onset of KHI \citep{Soler_2012AA...546A..82S}. However, we consider this effect not significant due to the very rapid timescales of these mechanisms. 

Another limiting factor of our model is the symmetry, especially in the $y$-direction, based on the spatial properties of the ideal kink mode. This assumption implies that no changes in polarisation nor coupling between azimuthal wave modes take place, and that therefore most of the energy will remain in the $m=1$ kink mode polarised along the $x-$direction. Density inhomogeneities, like those produced from the KHI in our model, can lead to mode coupling. Such effects, which we consider secondary, may lead to detectable Doppler asymmetries between diametrically opposite regions in the flux tube, and should be studied in more detail in the future.

The symmetry of our model implies the absence of twist, which may probably be the most limiting factor, but only on the KHI. A small amount of twist may inhibit the instability in the linear regime \citep{Soler_2010ApJ...712..875S}. However, the presence and degree of twist in a prominence is a subject of debate. Also, no numerical parameter study exists investigating the full effect of twist in a more realistic model. On the other hand, a transverse MHD wave still produces multiple current sheets in a flux tube with full coherent twist \cite{Ofman_2009ApJ...694..502O}. In our simulations the KHI produces partial twist of the flux tube, which slows down the instability itself but does not suppress it.

\section{Forward modeling}\label{forward}

\subsection{Conventions on LOS angle and spatial resolution}\label{convention}

For the forward modeling we define the LOS angle such that $0^{\circ}$ is parallel to the positive y-axis, and $90^{\circ}$ is parallel to the positive $x$-axis, which is the axis of oscillation (see Fig.~\ref{fig_init}).

For correct comparison with observations with a given instrument with resolving power of X we degrade the original spatial resolution of the numerical model by first convolving the image of interest with a Gaussian with FWHM of X, simulating a PSF of roughly the same size. For each pixel we then sum over a region of X in size centered on the pixel.

When comparing with observations it is important to keep in mind that the spatial resolution of SOT (0.2\arcsec) is approximately twice as good as that of IRIS ($0.33\arcsec - 0.4\arcsec$). Accordingly, the imaging and spectroscopic results presented here have been convolved with different Gaussian functions whose FWHM ratio is in accordance to the PSF ratio between SOT and IRIS. This is especially important when predicting the emission in slits that are relatively far away from the observed threads, as in the present case.

\subsection{Optically thick approximation}\label{OT_approx}

Radiative transfer calculations have been performed for synthesis of the \ion{Mg}{2}~h\&k and \ion{Ca}{2}~H\&K resonance lines in non-LTE (see Paper~1). We used the two-dimensional version of the RH code \citep{Uitenbroek_2001ApJ...557..389U}, based on the method described in \citet{Rybicki_1992AA...262..209R}. Results with RH indicate that only rays crossing the prominence core are optically thick and that only the boundary layer contributes to the emerging intensity. Furthermore, the intensity in the boundary layer is optically thin, and therefore mostly temperature dependent. Since RH is computationally intense, we have adopted an approximation to optically thick emission with the FoMo code \citep[https://wiki.esat.kuleuven.be/FoMo,][]{Antolin_VanDoorsselaere_2013AA...555A..74A}, which allows to perform forward modeling of time dependent results for the \ion{Mg}{2}~k line for the entire numerical box. This approximation is based on the RH results described in Paper~1 and consists solely of limiting the integration path of the emission line flux ($G_{\lambda}(T,n)n_e^{2}$ function) along a given LOS ($T$ and $n_e$ denote, respectively, the temperature and electron number density at a given position along the LOS), as follows. At a given time $t$ during the simulation, the specific intensity at wavelength $\lambda$ along a given LOS is proportional to:

\begin{equation}
I_{\lambda,ray}^{\mbox{OT}}\propto\int_0^{l}G_{\lambda}(T,n_e)n_e^{2}dl^{\prime},
\end{equation}
$l$ such that 
\begin{equation}
\int_0^{l}G_{\lambda}(T(t),n_e(t))n_e(t)^{2}dl^{\prime}=\mbox{OT}\times\int_0^{L_{max}}G_{\lambda}(T(0),n_e(0))n_e(0)^{2}dl^{\prime}
\end{equation}
where $L_{max}$ is the maximum path length along the LOS and OT is a parameter between 0 and 1 indicating the degree of optical thickness. OT$ = 1$ is the optically thin result\footnote{this is the case for $t=0$, and is a sufficiently good approximation for $t>0$}. The left hand side is the integral up to a distance $l$ along the LOS of the emission line flux (and therefore part of the total intensity). The limit of the integral, $l$, is physically similar to the distance along the LOS where the optical thickness at the wavelength $\lambda$ becomes unity. The right hand side of the equation is the optically thin intensity at time $t=0$ (disregarding the abundance of the element) multiplied by the factor OT. Fixing the parameter OT defines $l$, and therefore the unknown opacity of the material is determined. We therefore assume that the optical thickness is roughly inversely proportional to the optically thin intensity of the material. As a proportionality factor we assume that the maximum achievable intensity along such LOS rays is roughly the optically thin intensity at time $t=0$ (right hand side with OT$ =1$). This is reasonable since the model stays mostly in the linear regime (no major longitudinal perturbation) which leads to little mass displacement along the flux tube axis. For instance, with a value of OT$=0.5$ for a LOS crossing the center of the flux tube only rays starting from the center contribute to the emerging intensity at time $t=0$.

In optically thin conditions the emission line flux in the \ion{Mg}{2}~k line obtained from CHIANTI has a peak at $\log(T) \approx 4.2$, leading to a ring shaped structure similar to that of the source function obtained from RH (see Fig.~8 in Paper~1). This optically thick approximation therefore recovers the essential aspects of the radiative transfer calculations and is justified by the mainly temperature dependent source function in \ion{Mg}{2}~k and \ion{Ca}{2}~H. 

\subsection{Thread-like structure, out-of-phase behaviour, edge broadening and heating}

\begin{figure}[!ht]
\begin{center}
$\begin{array}{cc}
\includegraphics[scale=0.5]{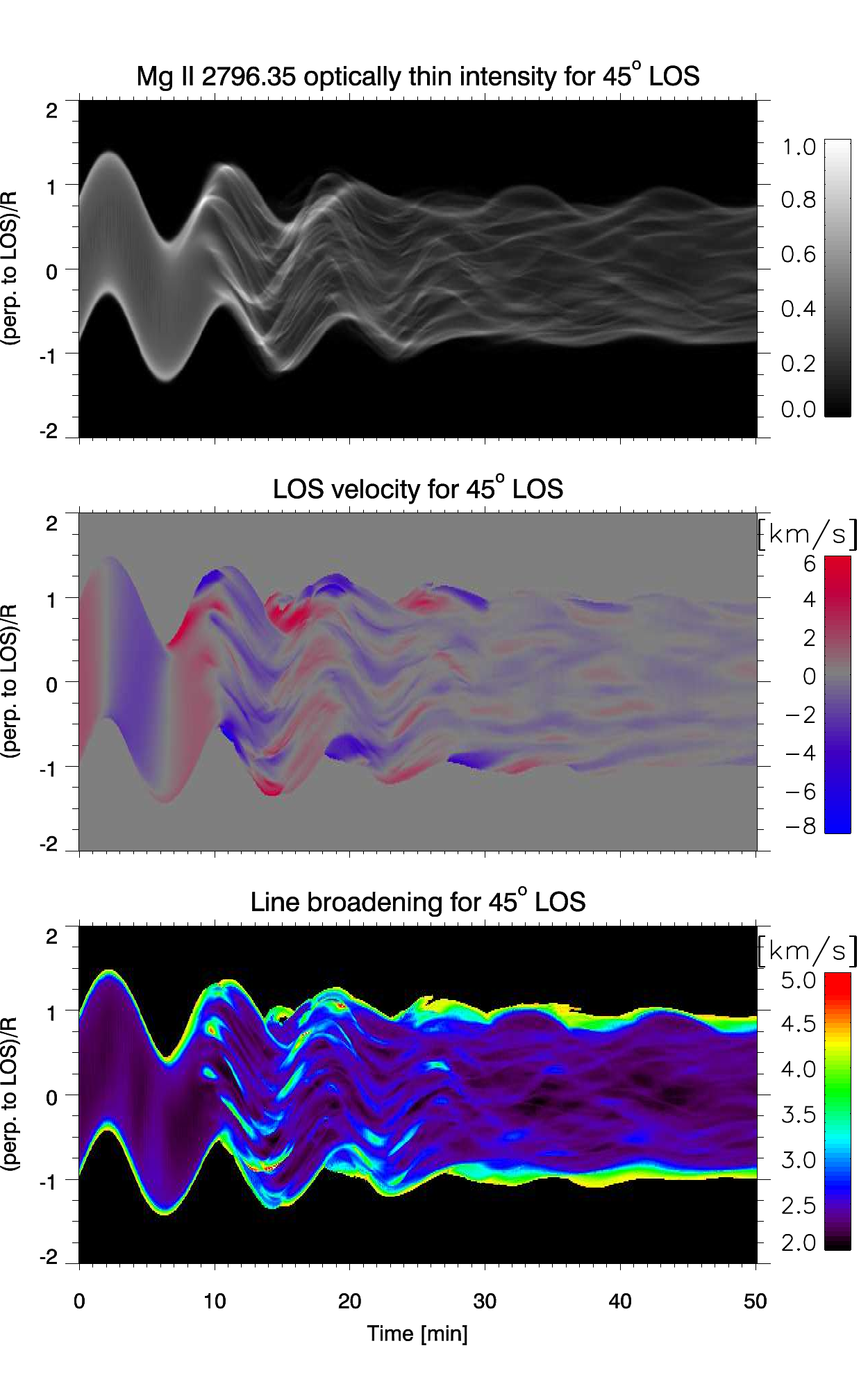} & 
\includegraphics[scale=0.5]{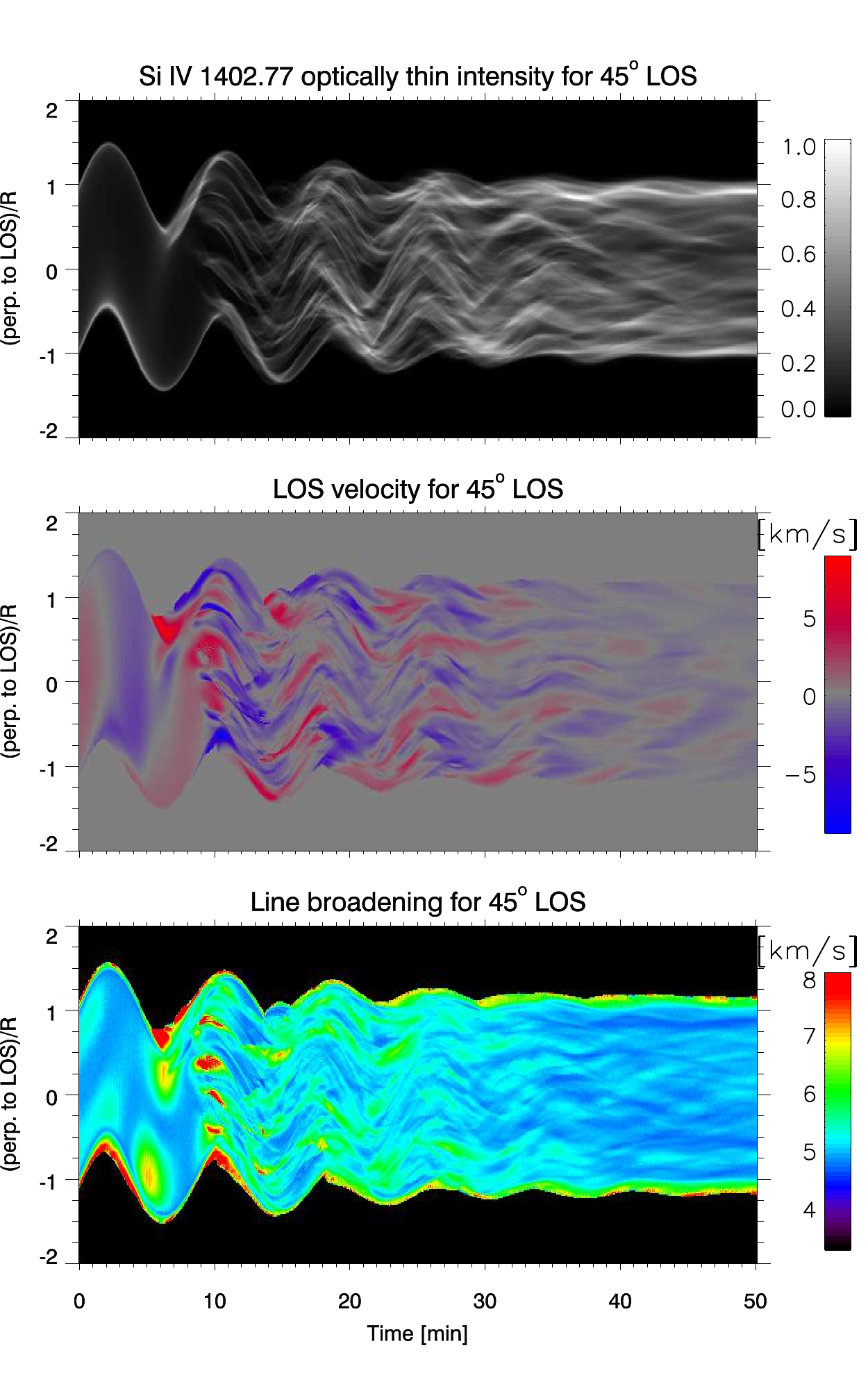}
\end{array}$
\caption{From top to bottom we show time slices of the optically thin intensity, LOS velocity and line broadening for the \ion{Mg}{2}~k line (left panels) and the \ion{Si}{4}~1402.77~\AA\, line (right panels) that a theoretical slit would capture being perpendicular to the prominence axis, located at the apex and making a LOS angle of 45$^{\circ}$ (see section~\ref{convention}). We further assume that the slit spatial resolution is equal to the numerical resolution. LOS velocity and line broadening are calculated with single Gaussian fitting to the line profiles. 
\label{fig_idw}}
\end{center}
\end{figure}

In the top left panel of Fig.~\ref{fig_idw} we show the optically thin case of the \ion{Mg}{2}~k intensity for a theoretical slit located perpendicular to the flux tube, at the apex, and for a LOS angle of $45^{\circ}$. Here we further assume that the slit has a spatial resolution equal to that of the numerical resolution. As for the coronal case \citep{Antolin_2014ApJ...787L..22A}, thread-like structure rapidly appears after just one period of oscillation in the prominence case and permeates the entire flux tube. The threads correspond to the KHI vortices, which, as explained in section~\ref{KHI} initially occur at the surface of the flux tube boundary and rapidly degenerate in turbulent-like flows drifting towards the core of the prominence. 

From the synthetic emission profiles obtained with FoMo we calculate the moments by single Gaussian fits to the line profiles (double Gaussian fits are not necessary due to the relatively low velocity amplitudes). The middle and lower left panels of Fig.~\ref{fig_idw} correspond, respectively, to the LOS velocity and line broadening obtained in this way from the previously mentioned theoretical slit. In the Doppler velocity panel, apart from the initial kink displacing the flux tube in the POS, the onset of resonant absorption can be seen at the edges right from the beginning as enhanced Doppler signals, which become rapidly shifted in time due to phase mixing (more precisely, due to the change of the Alfv\'en speed across the boundary layer). This phase shift of the azimuthal flows at the flux tube boundary is ultimately responsible for the out-of-phase behaviour with the POS motion, as explained in Paper~1, which we further elaborate in the following sections. Resonant absorption is also visible in the line broadening panel, where clear enhancement is highly localized at the flux tube boundaries, where the azimuthal flows are directed along the LOS. 

As the KHI is triggered, at this high spatial resolution traces of the KHI vortices can be distinguished in the $x-t$ diagrams of the Doppler velocity and line broadening like thread-like structure similar to that in the intensity panel. This reflects the fact that the vortices, while embedded in the large-scale flow generated by the kink wave, have their own characteristic small-scale vortex flows. The KHI extracts the energy from the resonant layer and dumps it into heat through ohmic and viscous dissipation at the generated current sheets and vortices. A result of this mechanism is the generation of turbulence as the energy of the large eddies cascades down to lower eddies. This is clearly shown in movie~1 (see also movie~2 in Paper~1), where the higher resolution allows the generation of a higher number of small-scale current sheets and vortices, leading to a higher rate of dissipation. This results in an enlargement of the flux tube boundary, where significant heating occurs, and a corresponding enhancement and enlargement of the line broadening around the flux tube boundary, as shown in the lower left panel of Fig.~\ref{fig_idw}. Importantly, this enlargement is absent in the \ion{Mg}{2}~k intensity panel but is present in the Doppler velocity maps. This effect will further be analysed in section~\ref{angle}.

The heating can also be clearly observed in synthetic \ion{Si}{4} images, as shown in the right column panels of Fig.~\ref{fig_idw}. This column shows the same quantities as in the left column but for \ion{Si}{4}~1402.77~\AA, which has a formation temperature around $80,000~$K. As the flux tube with its strand-like structure fades out in \ion{Mg}{2} (top left panel), similar structure fades in in \ion{Si}{4} (top right panel). The spectral signatures shown in the middle and lower right panels (Doppler velocity and line broadening, respectively) have essentially the same features as for \ion{Mg}{2}. However, it is worth noting that the fine-scale structure in intensity, Doppler and line width is more clearly seen in the TR line at this high resolution, since the temperature formation of this spectral line captures better the heating events (current sheets and vortices) occurring around the boundary. 

\subsection{Effect of optical thickness}\label{optic}

\begin{figure}[!ht]
\begin{center}
$\begin{array}{c}
\includegraphics[scale=0.4]{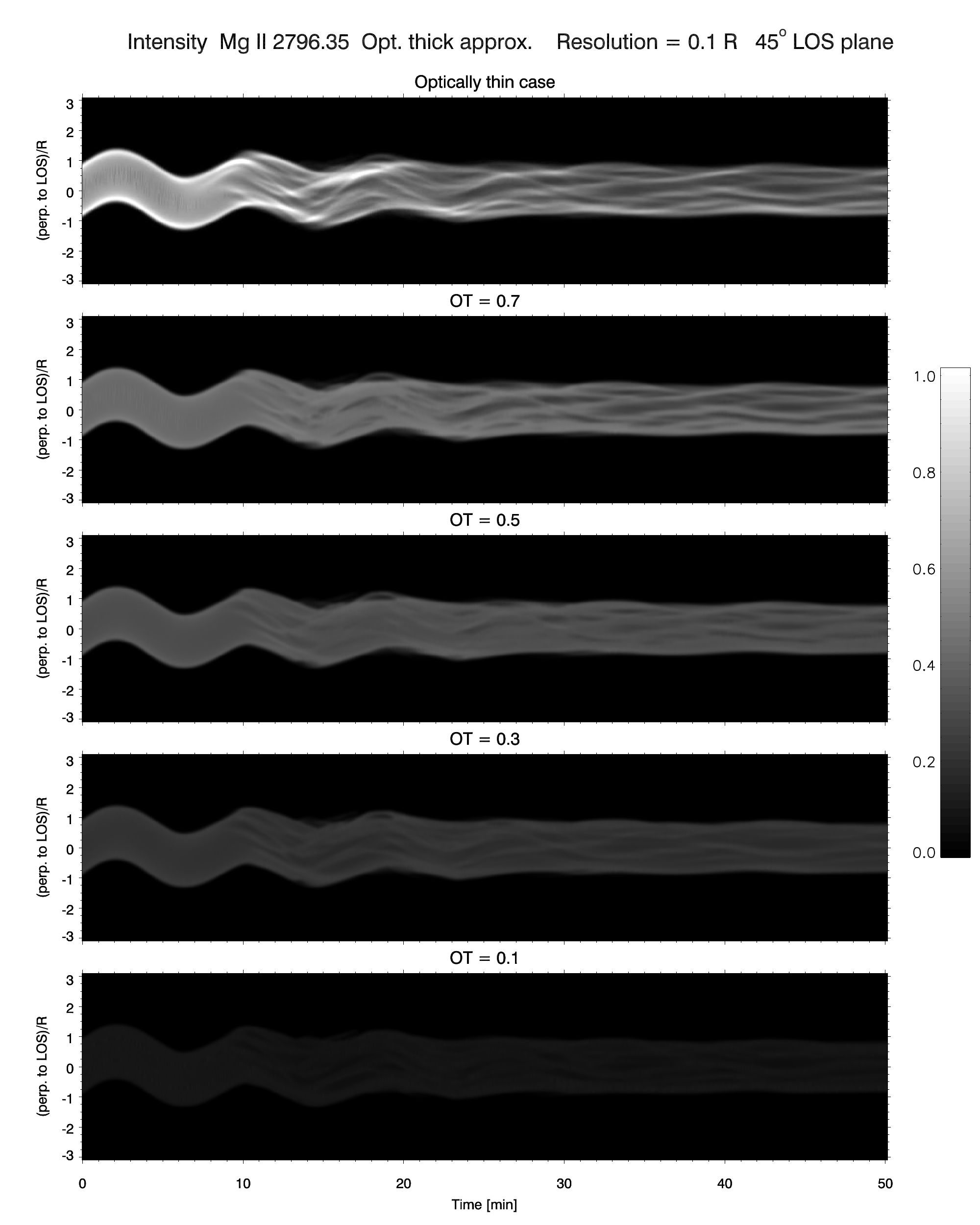}
\end{array}$
\caption{Effect of the optically thick approximation explained in section~\ref{OT_approx} on a time slice of the emerging \ion{Mg}{2}~k intensity for a slit perpendicular to the prominence axis, located at the apex and making a LOS angle of $45^{\circ}$. From top to bottom we show increasing degrees of the optically thick parameter OT, from purely optically thin (OT$=1$) to highly optically thick (OT$=0.1$). We show the effect for a spatial resolution of $0.1~R$.
\label{fig_OT}}
\end{center}
\end{figure}

In the simple geometry considered of a single flux tube we show in Fig.~7 of Paper~1 that chromospheric emission in \ion{Mg}{2}~h\&k and \ion{Ca}{2}~H\&K comes from the most dynamically important (ring-shaped) section of the flux tube, where resonant absorption and dynamic instabilities such as KHI take place. The ring shape of the source function in \ion{Mg}{2}~k and \ion{Ca}{2}~H are a temperature effect from the optically thin layer of material surrounding the prominence core. These results justify the approximation to the \ion{Mg}{2}~h\&k intensity based on optically thin calculations with the FoMo code explained in section~\ref{OT_approx}. The result of this approximation is shown in Fig.~\ref{fig_OT}, where the emerging intensity in \ion{Mg}{2}~k for several values of the OT parameter is plotted with respect to time for a theoretical slit located at the flux tube apex, perpendicular to the axis and for a LOS of 45$^{\circ}$ and a spatial resolution of $0.1~R$. The decrease in intensity with decreasing value of the OT parameter is due to the lower column density contributing to the emerging LOS ray. For instance, the case of $OT=0.5$ corresponds to an optically thick case in which the radiation along the center of the tube crosses roughly half the tube only. Several values of the OT parameter have been tried, from 0.1 to 0.7. We can see that thread-like structure is always obtained (except the case OT$=0.1$, which is too dark relative to the optically thin case), although the number of threads increases for decreasing optical thickness, as expected.

\begin{figure}[!ht]
\begin{center}
$\begin{array}{c}
\includegraphics[scale=0.5]{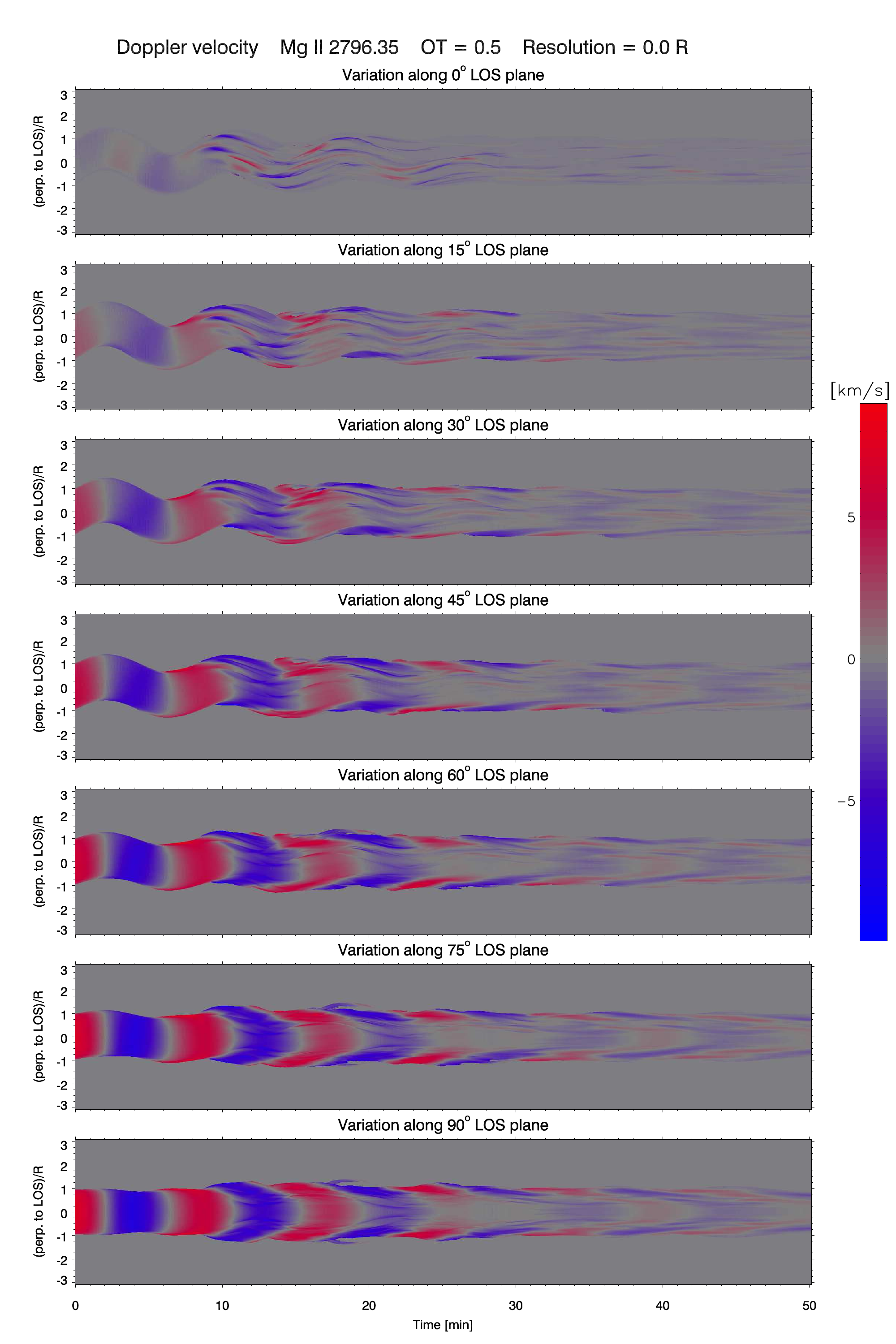}
\end{array}$
\caption{Time slice of the Doppler velocity for \ion{Mg}{2}~k along a slit perpendicular to the prominence axis, located at the apex. The parameter OT is set to 0.5, corresponding to a moderately optically thick case. The spatial resolution is the same as the numerical resolution. Each panel corresponds to a different LOS angle.
\label{fig_doppler}}
\end{center}
\end{figure}

In Fig.~\ref{fig_doppler} we show the LOS velocities that result for the optically thick case OT$ = 0.5$ of \ion{Mg}{2}~k. The figure corresponds to the time slice from a theoretical slit located at the center of the tube, perpendicular to its axis, for different LOS angles. The effect of optical thickness on LOS velocity is only significant when observing at shallow LOS angles. In other words, optical thickness in general does not affect the measured LOS velocity. Indeed, due to the symmetry of the kink mode, only at low angles ($<15^{\circ}$) the Doppler signals from opposite regions with respect to the loop's center plane ($x$-axis) tend to cancel each other. If the loop's core is opaque in a given wavelength only one side of the loop's surface will contribute, avoiding cancellation. In Fig.~\ref{fig_doppler} Doppler signals can be seen even for a $0^{\circ}$ angle.

\subsection{Effect of LOS angle}\label{angle}

\begin{figure}[!ht]
\begin{center}
$\begin{array}{c}
\includegraphics[scale=0.5]{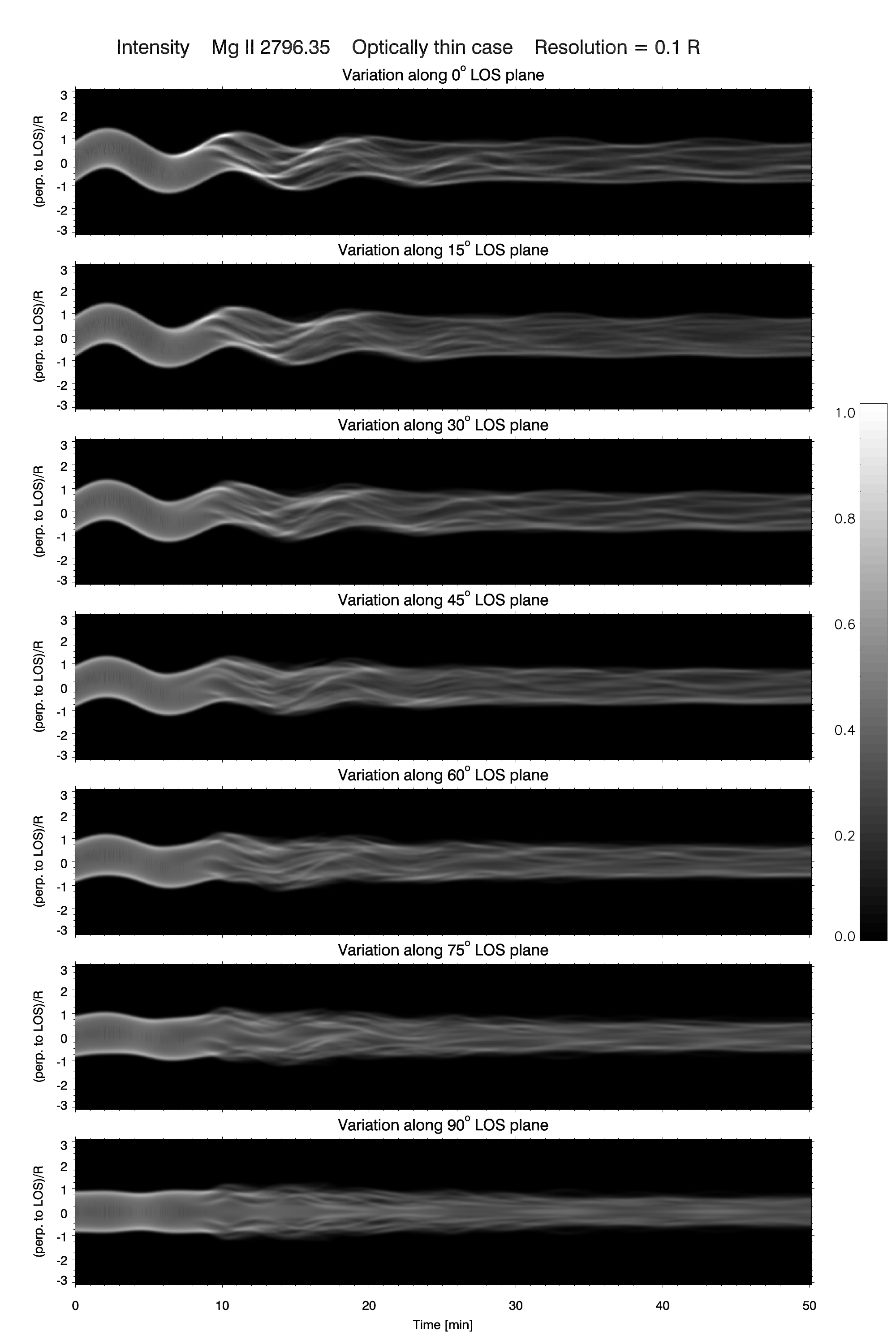}
\end{array}$
\caption{Time slice of the optically thin intensity for \ion{Mg}{2}~k  along a slit perpendicular to the prominence axis, located at the apex. The spatial resolution is $0.1~R$. Each panel corresponds to a different LOS angle.
\label{fig_ang_mg2}}
\end{center}
\end{figure}

In Fig.~\ref{fig_ang_mg2} we show the effect of LOS angle on the emerging intensity in \ion{Mg}{2}~k for the case of optically thin emission and a spatial resolution of $0.1~R$. As seen in the figure, the thread-like structure appears at any angle. Due to LOS effects, apparent crossing of threads is observed, especially for the optically thin case. This corresponds in the model to either twisting within vortices or to the combination of several vortices along the LOS. As in the coronal case the number of threads and the distribution of widths is highly dependent on the initial kink amplitude and the width of the boundary layer. Similarly, since vortices move coherently (they are embedded in the large scale azimuthal flow) threads detected at coarse spatial resolution actually correspond to groups of threads, as can be seen comparing Figs.~\ref{fig_idw} and \ref{fig_ang_mg2}. The effect of spatial resolution will further be discussed in section~\ref{spatial}.

The preferential occurrence of resonant absorption and KHI vortices along the direction of oscillation (with azimuthal flows preferentially directed along the $x$-axis) generates an asymmetry in the flux tube cross-sections for all physical quantities, making these more elliptic. Indeed, the generation of current sheets and vortices where most of the heating takes place is concentrated in lobes around the prominence core, with maxima at $\pm90^{\circ}$ from the $x$-axis and minima along the $x$-axis (clearly seen in movie~1). These lobes are expected from the spatial distribution of the $m=1$ mode \citep[as shown also by][; see their Fig.~4]{Pascoe_etal_2011ApJ...731...73P}. A corresponding decrease of density and increase of temperature is obtained along these lobes, which broadens inwardly the flux tube boundaries. The emission line flux in \ion{Mg}{2}~k is thus diminished, partially due to the density squared dependence of the intensity, which is expected under the optically thin conditions of the boundary layer (contributes in generating sharp threads), but especially to the increase in temperature in the same layer, which brings the plasma gradually out of the temperature formation of \ion{Mg}{2}~k. Accordingly, this effect leads to a thinning of the flux tube in time for most LOS angles, as shown in Figs.~\ref{fig_idw} and \ref{fig_ang_mg2}. Also, a difference in width of the flux tube is readily seen when observing at different LOS angles, as shown in Fig.~\ref{fig_ang_mg2}. On the other hand, the thinning is not observed in synthetic \ion{Si}{4} intensity images, as shown by Figs.~\ref{fig_idw} and \ref{fig_ang_si4}, since the temperature of the heated plasma enters the range of temperature formation of this spectral line. This LOS angle effect has also been reported in the coronal case \citep{Antolin_2014ApJ...787L..22A}.

Through cross-correlation analysis we find that the lifetime of a KHI vortex is roughly one period. However, as in the coronal case \citep{Antolin_2014ApJ...787L..22A} it is difficult to estimate precisely in the model the lifetime of threads, due to the continuous generation and destruction of vortices and the projection effects. A new vortex can appear at the location of a previously destroyed vortex, thus producing apparently longer lived thread-like structure. This can be checked in Fig.~\ref{fig_ang_mg2} and movie~1 (especially clear in the high resolution run). Furthermore, we can expect such vortices to be longer lived in the real corona since turbulence is only independent on viscosity for larger viscous Reynolds numbers.

The effect of the LOS angle on the LOS velocity is shown in Fig.~\ref{fig_doppler}, where the optically thick case of OT$ = 0.5$ is shown for \ion{Mg}{2}~k. The time slice clearly shows that the red-blue patterns at the edges of the tube are shifted towards $90^{\circ}$ out of phase with respect to the Doppler shifts at the center, and $180^{\circ}$ out of phase with respect to the transverse tube displacement in the POS. This results in characteristic arrow (“$>$”) shaped structures in such time slices, which can be observed for basically any LOS angle and any optical thickness (even at shallow angles as long as the line opacity is significant, as explained before). This characteristic out-of-phase behavior from resonant absorption can be clearly seen after just half an oscillation period (before the onset of the KHI), and remains regardless of the presence or absence of the instability. However, the detectability of this out-of-phase behavior and the instability is significantly increased because of the spatial broadening of the boundary layer that the instability produces.

\begin{figure}[!ht]
\begin{center}
$\begin{array}{c}
\includegraphics[scale=0.5]{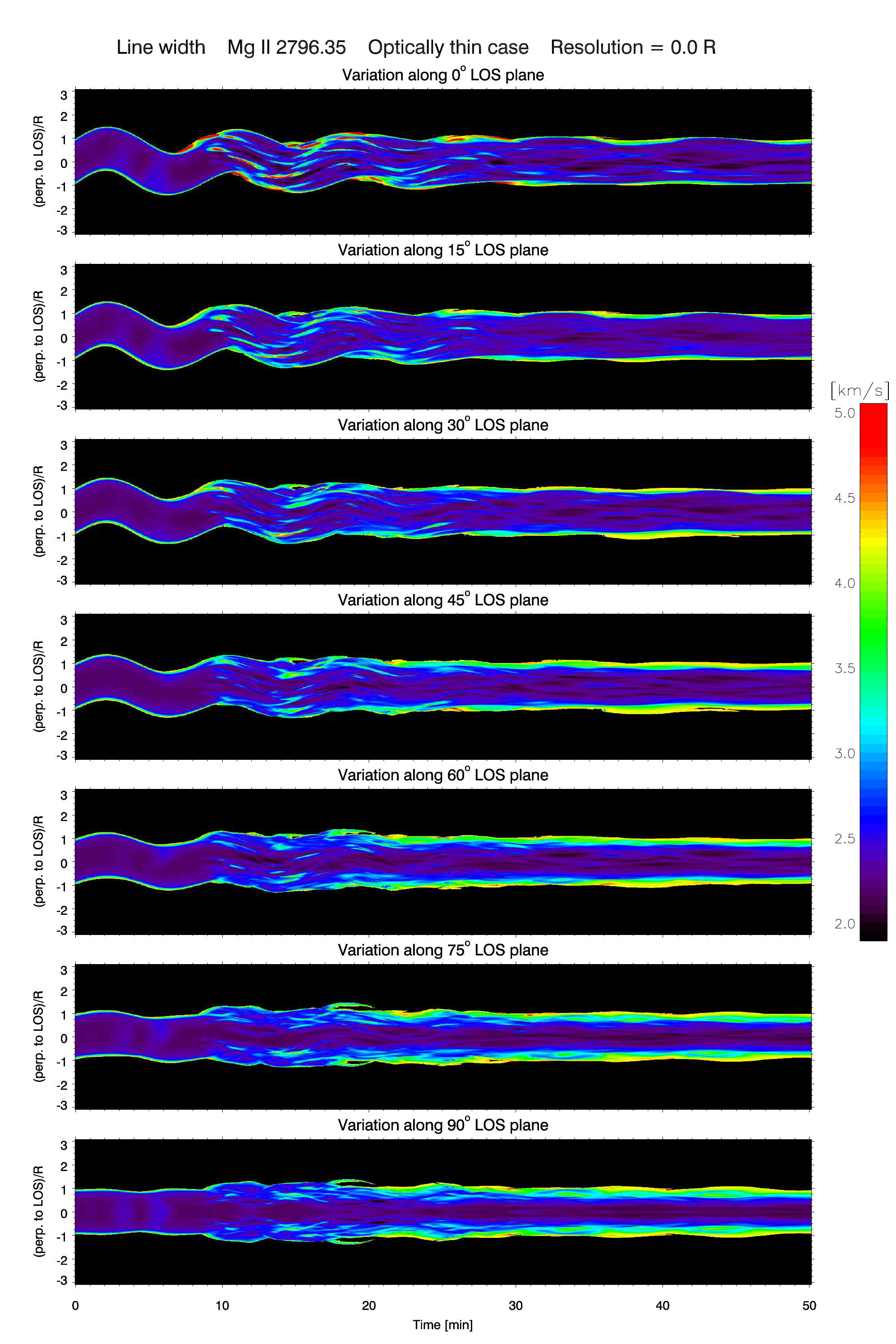}
\end{array}$
\caption{Time slice of the line broadening in \ion{Mg}{2}~k for the case of optically thin intensity along a slit perpendicular to the prominence axis, located at the apex. The line broadening is calculated from single Gaussian fits and corresponds to the FWHM of the spectral line. The spatial resolution is $0.1~R$. Each panel corresponds to a different LOS angle.
\label{fig_wlin}}
\end{center}
\end{figure}

\begin{figure}[!ht]
\begin{center}
$\begin{array}{c}
\includegraphics[scale=0.5]{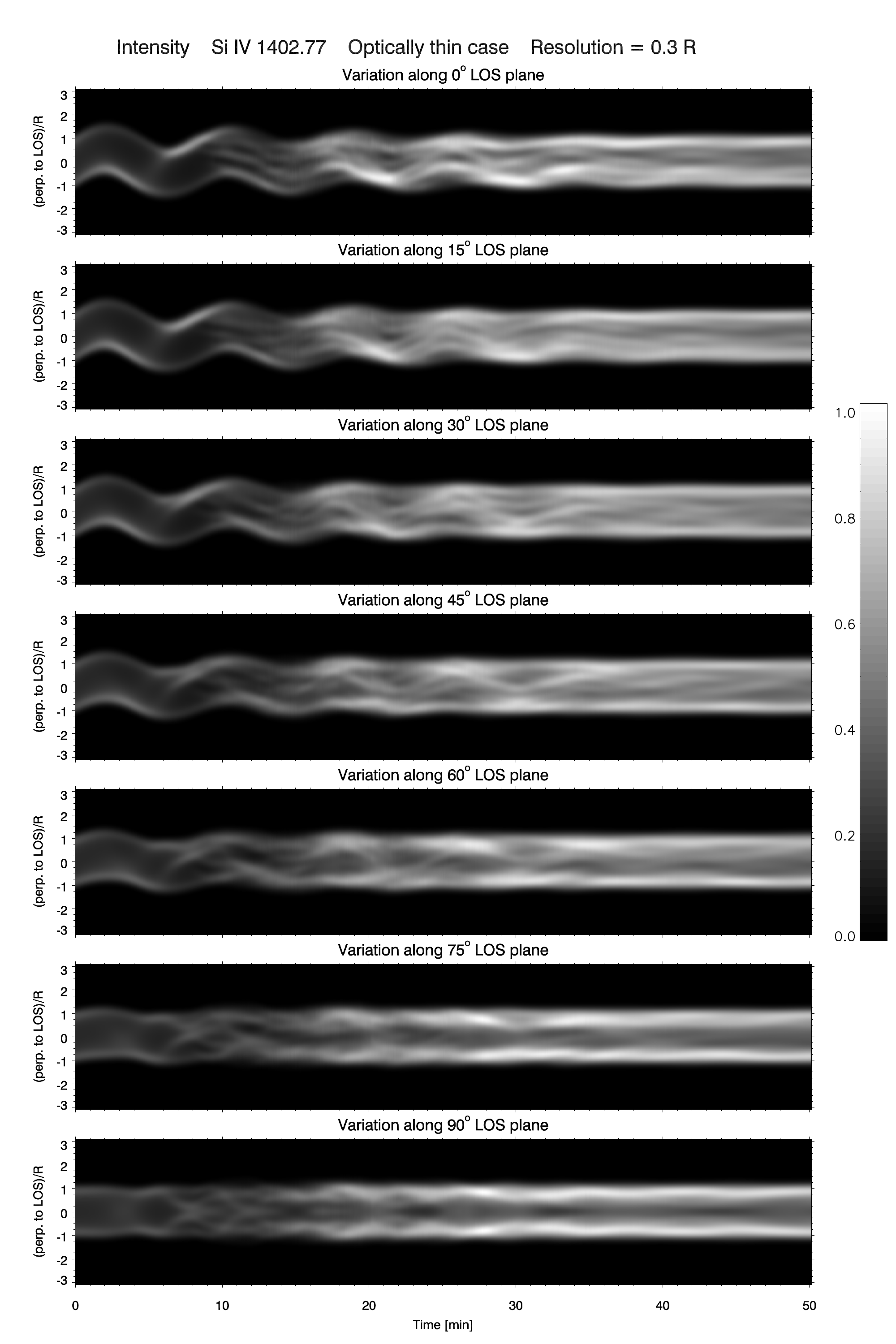}
\end{array}$
\caption{Time slice of the intensity for \ion{Si}{4} along a slit perpendicular to the prominence axis, located at the apex. The spatial resolution of 0.3 R. Each panel corresponds to a different LOS angle.
\label{fig_ang_si4}}
\end{center}
\end{figure}

In Fig.~\ref{fig_wlin} we show the effect of LOS angle on the measured line broadening of the \ion{Mg}{2}~k line. As for the LOS velocity, the maximum values are found around the boundary layer, highly localized at first in the resonant layer before the occurrence of the KHI, and more widespread after occurrence. Part of this line broadening is not due to resonant absorption but to the fact that at the edges the LOS crosses more material with different velocities (especially in optically thin conditions), which can clearly be seen at $t=0$ for a $0^{\circ}$ LOS (angle for which the azimuthal component is largely absent at the edges). This effect has also been described by \citet{VanDoorsselaere_2008ASPC..397...58D}. As seen in the figure, line broadening becomes more detectable (occupies an apparently larger area of the flux tube) when observing along the axis of oscillation. This is expected since such LOS rays cross the lobes of oscillation of the $m=1$ mode. The maximum values, however, are detected at shallow angles, just after KHI occurrence. This is because LOS rays at such angles cross vortex flows with different vorticity occurring next to each other. The flow within a given vortex introduces strong velocity shear, especially when embedded in the azimuthal flow amplified by the resonance. Further small vortices are then generated on top of the initial vortex, with different small-scale flow direction (different vorticity) but still embedded in the larger azimuthal flow. This can be clearly seen in the vorticity panels of Fig.~\ref{fig_model} and in movie~1 (see also movie~2 in Paper~1). These results could potentially explain observations of line broadening enhancement at loop edges or in inter-loop planes  \citep{Doschek_etal_2007ApJ...667L.109D}.

\section{Discussion}\label{discussion}

\subsection{Intensity variability}

As discussed in Paper~1 two kinds of intensity variability can be noticed in the observations of prominence threads. The first kind is linked to flows or waves in the threads and occurs on short timescales of one oscillation period or less. The second kind is linked to the dimming in \ion{Ca}{2}~H intensities and corresponding intensity increase in \ion{Si}{4} and is therefore linked to heating. The latter is more gradual and occurs in timescales of $1-2$ periods. 

As in the observations, the intensity variability obtained in our simulations has also two components: a short timescale component linked to dynamics and a slightly longer component linked to large scale heating. These two components can be seen in Fig.~\ref{fig_idw} (also Figs.~\ref{fig_ang_mg2} and \ref{fig_ang_si4} for different angles and spatial resolutions). The former can be seen within the flux tube (and thus only at high spatial resolution) and is associated to the lifetime of KHI vortices, on the order of one period or less. The latter component is associated to the heating of the entire flux tube and occurs in a timescale of $10-15$ min (roughly two periods, between times $t=10$ and $t=25$ min), and a few minutes shorter in the case of the high resolution run.

Although a proper parameter space study is needed for investigating the full effects of the initial transverse wave amplitude, results from the previous study \citep{Antolin_2014ApJ...787L..22A} indicate that an increase (decrease) of amplitude (within the linear regime) produces stronger (weaker) current sheets and larger (smaller) vortices, due to the increase (decrease) in velocity shear. Stronger (weaker) ohmic and viscous dissipation and turbulence is thus obtained. We can then expect that chromospheric plasma within the small-scale vortices composing the turbulence will tend to be more or less rapidly heated depending on the amplitude of the oscillation. Specifically, smaller amplitudes produces longer lived flux tubes (as observed in chromospheric lines). This model can therefore explain the observed tendency of lifetime decrease with height, shown in Fig.~2 of Paper~1, since amplitude of the transverse motion is observed to increase with height. It is also likely that other longitudinal effects besides heating, such as the length of the flux tube or the mass flows, play an important role in the morphology and lifetimes of threads \citep{Zhou_2014RAA....14..581Z}.

\subsection{Damping of transverse oscillations}\label{damping}

Although signatures of damping can be seen in the threads, constant amplitude oscillations and even sudden increases can also be seen in movies~$3-6$ of Paper~1. Such features may seem at first contradictory with the resonant absorption scenario, since this one indicates that a net transfer of energy should take place from the global kink mode (affecting the transverse POS motion) to the local Alfv\'en waves in the resonant layer (affecting the azimuthal flow and therefore the LOS velocity component). However, there are a number of realistic scenarios that indicate that damping is not a necessary observational characteristic of resonant absorption. For instance, a decrease in the density of the flux tube or the background can lead to an increase of the amplitude of the oscillation, thus counteracting the damping from resonant absorption (in other words, an increase in amplitude does not imply an increase of energy). Similarly, a decrease of the flux tube's density or an increase of the background's density increases the damping time. A decrease in density would also lead to a decrease of the opacity of the material, making the oscillation hard to follow. Another possibility for non-decreasing kink amplitudes is one in which the oscillation energy is ceaselessly injected into a damped harmonic system through a non-resonant continuously operating external force. This has been suggested by \citet{Anfinogentov_2013AA...560A.107A} based on undamped transverse oscillations in loops observed with AIA. Such external force could be readily (and is most likely) achieved by continuous convective motions at the photosphere. In this scenario, the energy that is continuously dumped into the system is damped by the resonant absorption (that we observe) and turned into heat, without actually changing the amplitude of the wave significantly. Another possibility for the apparent decay-less oscillations is provided by the numerical model and suggests that the observed threads may correspond to the obtained threads in the simulation, which are the result of KHI vortices. The observed POS oscillations would then correspond to those of the KHI vortices, which, as explained in section~\ref{spatial}, produce oscillating thread-like structure that damps, as an ensemble, on longer timescales. 

\subsection{$\lambda-y$ symmetry}

\begin{figure*}[!ht]
\begin{center}
$\begin{array}{c}
\includegraphics[scale=0.5]{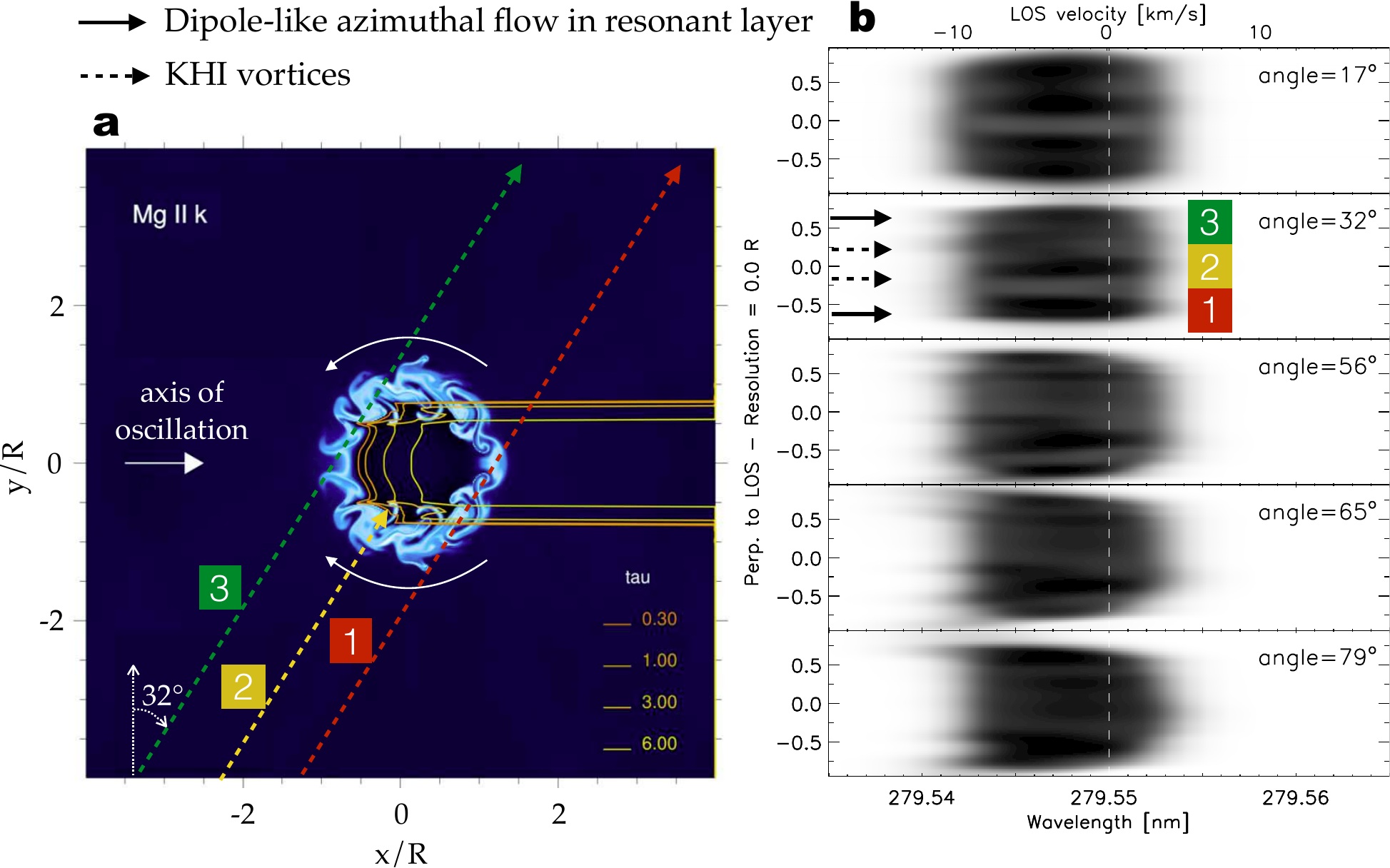}
\end{array}$
\caption{Doppler shifts variation across the flux tube in the \ion{Mg}{2}~k line. $\lambda-y$ diagrams (\textit{panel b}) produced with RH in the \ion{Mg}{2}~k line from a slit placed at the center and perpendicular to the tube's axis for several inclination angles (in negative colors). The diagrams are made from the same snapshot as in Fig.~8 of Paper~1 (shown here in \textit{panel a}, rotated for comparison with Fig.~\ref{fig_model}). The optical thickness values (yellow contours) are reproduced from that figure. The curved white arrows on \textit{panel a} indicate the direction of the large scale azimuthal flow at that instant in time. Three rays denoted as red (1), yellow (2) and green (3) denote 3 examples of LOS rays going through the flux tube at an inclination of $32^{\circ}$. The Doppler shifts at these 3 locations along the slit are marked on \textit{panel b} in the corresponding panel of that angle. Solid and dashed arrows indicate Doppler shift features generated by the resonant flow and the KHI vortices, respectively. Except for the individual Doppler signals from the KHI vortices (which can have a sporadic different behavior) the Doppler signal across the flux tube is generally symmetric with respect to a LOS through the center of the flux tube: the leading and trailing edges (rays 1 and 3) exhibit a similar Doppler signal.
\label{fig_lambda}}
\end{center}
\end{figure*}

The characteristic arrow shaped structure seen in the Doppler maps of Fig.~\ref{fig_doppler} implies that the resulting $\lambda-y$ profile is generally symmetric, where $y$ denotes the distance across the flux tube in the POS. Indeed, by taking a specific time in the figure and plotting the Doppler velocity versus perpendicular distance, the resulting profile is symmetric with respect to the center of the flux tube. Such symmetry has strong consequences since it implies that in an observational setup such as that in Paper~1 (Fig.~5) a coherent signal between the slits is expected. To better understand why this is so, we have produced Fig.~\ref{fig_lambda} with RH from a random snapshot of our simulation (same as in Fig.~7 of Paper~1). Similar to Fig.~\ref{fig_doppler}, the figure shows that when viewing along an inclined ray that has a significant perpendicular component with respect to the direction of oscillation, the Doppler signal is to a large extent symmetric across the flux tube (as projected onto the plane of the sky). In the fixed $x-y$ frame of reference, in which the kink mode oscillation occurs along the $x$-direction, the $y$ components of the azimuthal flows have different signs (thin white arrows) and become zero at the trailing and leading (with respect to the direction of kink oscillation, i.e., the $x$-direction) edge of the flux tube. For our spectral lines (\ion{Mg}{2}~k and \ion{Ca}{2}~H), rays~1 and 3 which cross the edges of the flux tube will go through only optically thin material, as depicted in Fig.~\ref{fig_lambda}. For ray~3 and basically any viewing angle there is an imbalance of red-shifted and blue-shifted material resulting in a blueward Doppler shift. This imbalance is caused by the fact that the azimuthal flows are zero at the trailing and leading edges of the flux tube. A similar imbalance in ray 1 causes the same type of blueward Doppler shift. This explains the general symmetry of the velocity across the flux tube observed in Fig.~5 of Paper~1. There are small deviations to this symmetry (e.g., the jagged nature of the $\lambda-y$ plot between rays~2 and 3 in the right panel of Fig.~\ref{fig_lambda}). These deviations are caused by the fine-scale velocity signals associated with the KHI vortices on the boundaries of the flux tube. This is similar to what we see in the observations (Fig.~5 in Paper~1). We note that in Fig.~\ref{fig_lambda} the optical thickness of the flux tube in \ion{Mg}{2}~k is automatically taken into account and is thus not a free parameter. This is important for the comparison of our model results with the observations.

Only if the tube were completely optically thick, for very small shallow angles and in the case of a flux tube whose internal flows were resolved with the spectrometer (case in which the flows from the individual KHI vortices can be resolved) the symmetry of Doppler signals across the tube could be significantly changed because then only the side of the flux tube that faces the observer (and potentially the individual KHI vortices) contributes to the observed signal. In addition, the presence of the KHI thickens the layer in which significant azimuthal flows occur, thus facilitating a more symmetric Doppler pattern across the flux tube. Both of these effects cause significant differences between our model and previous models who considered completely optically thick flux tubes and did not include the KHI \citep{Goossens_2014ApJ...788....9G}. These different results clearly show the importance of proper radiative transfer modeling in order to forward model numerical results.

\begin{figure*}[!ht]
\begin{center}
\includegraphics[scale=0.4]{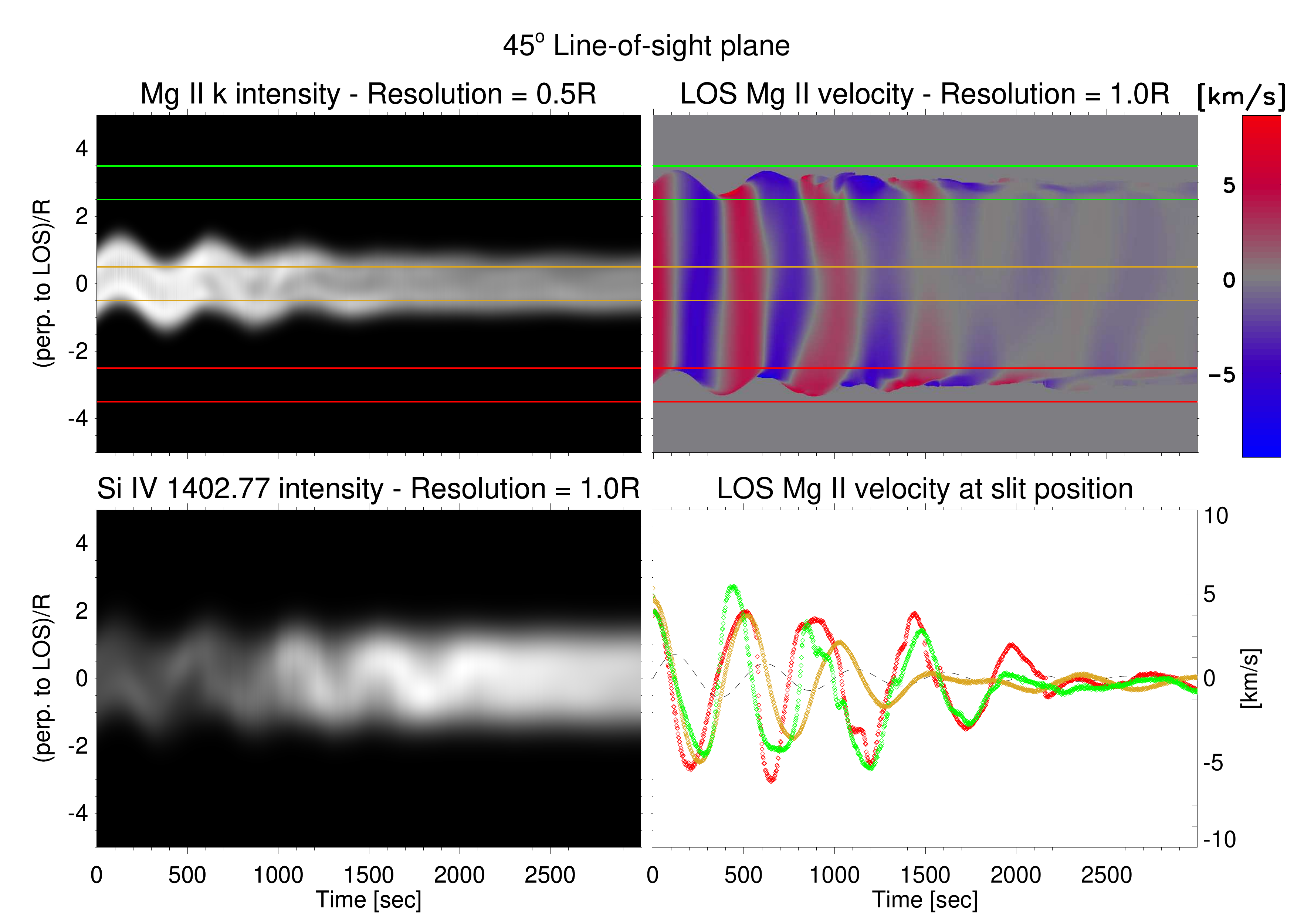}
\caption{Time slice of the optically thin intensity (\textit{top left panel}) and LOS velocity (\textit{top right panel}) in the \ion{Mg}{2}~k line along a slit perpendicular to the prominence axis, located at the apex and for a LOS angle of $45^{\circ}$. The corresponding time slice intensity in the \ion{Si}{4}~1402.77~\AA\, line is shown in the \textit{bottom left panel}. We set a spatial resolution of $0.5~R$ for the \ion{Mg}{2}~k intensity and $1.0~R$ for the LOS \ion{Mg}{2}~k velocity and \ion{Si}{4} intensity, simulating, respectively, what SOT and IRIS would observe in the case of an unresolved flux tube (here we assume the \ion{Ca}{2}~H intensity of SOT to be similar to the optically thin \ion{Mg}{2}~k intensity).
 In the \textit{bottom right panel} we show the Doppler signals along the green, orange and red slit locations, with respective colors (whose colored edges are shown in the top panels), centered at a distance of $3~R, 0~R$ and $-3~R$ from the flux tube's axis, respectively. The black dashed line corresponds to the transverse displacement of the flux tube in the POS, calculated from Gaussian fits to the intensity image for each time step. Times without Doppler signal correspond to regions for which the signal is too low.
\label{fig_coarse}}
\end{center}
\end{figure*}

\subsection{Dynamical coherence in the transverse MHD wave model}

As seen in Fig.~\ref{fig_doppler} and explained in sections~\ref{optic} and \ref{angle}, the thread-like structure and the characteristic arrow (“$>$”) shaped structures can be observed for basically any LOS angle and optical thickness in the present transverse MHD wave model. Such behavior is obtained by the combination of resonant absorption and the presence of KHI, and is predicted in the observations by our advanced radiative transfer calculations. As explained in the previous section, the obtained arrow shape in the Doppler maps means a $\lambda-y$ profile that is generally symmetric, where $y$ denotes the distance across the flux tube in the POS. This implies that in an observational setup such as that of Fig.~5 in Paper~1 a coherent signal between the slits is expected.

In a model with negligible resonant absorption, and therefore negligible azimuthal flow in the resonant layer, we would have the transverse displacement and the dipole-like azimuthal flow around the tube. Because of the little transfer of energy from the transverse motion into the resonant layer in such case, the plasma with azimuthal flows would mainly be external plasma, which we expect to be at a considerably different temperature (probably coronal) than the temperature within the flux tube. This would therefore be invisible to IRIS. However, as shown by Fig.~5 in Paper~1 the out-of-phase relation between the POS displacement and the LOS velocity exists over a significant distance across the threads, suggesting that the resonant layer is not bound to the thread but can be detected over a larger distance. This result therefore cannot be explained by a transverse MHD model without a resonant absorption layer (or one in which the resonant layer is highly confined to the boundary layer and is not broadened by any other mechanism). On the other hand, as shown by Figs.~\ref{fig_idw}, \ref{fig_doppler} and \ref{fig_ang_mg2} (see also Fig.~6 in Paper~1 and Fig.~\ref{fig_coarse}), in our numerical model combining resonant absorption and KHI the characteristic Doppler signal extends over a significant transverse distance with respect to the corresponding chromospheric intensity signal from the flux tube, matching the observations. 

Two effects contribute significantly to this spatial broadening. First, the KHI rapidly broadens (inwardly) the boundary layer after it sets in, clearly seen in Fig.~\ref{fig_model} and movie~1. The KHI extracts the energy out from the resonant layer and imparts momentum to the eddies around the boundary layer. As a result, the plasma layer with azimuthal flow has a significant width, which can then be detected with the spectrometer. As explained in section~\ref{angle} this spatial broadening is not observed at high resolution in the optically thin intensity images in \ion{Mg}{2} (with which actually a thinning of the flux tube is observed), but in those of \ion{Si}{4}, as shown by Figs.~\ref{fig_idw} and \ref{fig_ang_si4}, and in the Doppler and line broadening maps, as shown by Figs.~\ref{fig_idw} and \ref{fig_doppler}.

The second effect is perhaps even more significant and relies on instrumental aspects: when calculating intensities and velocities from our numerical model we have taken into account the different spatial resolutions of SOT and IRIS for correct comparison with the observations in Paper~1 (further explained in section~\ref{spatial}). The coarser resolution of IRIS ($0.33\arcsec - 0.4\arcsec$) with respect to SOT ($0.2\arcsec$) implies that the Doppler signal spreads over a wider distance than that delimited by the SOT threads, and corresponds at least to an area proportional to that of the IRIS PSF. Since the resolution of IRIS is roughly the same as the width of a thread (as shown in Fig.~5 of Paper~1), the signal should be detected at least a full diameter away from each side of the thread. This closely matches what we observe, as shown by Fig.~5 in Paper~1, thereby supporting the fact that our Doppler measurements with IRIS in Paper~1 actually are set by the resonant absorption region (and not the exterior).

It is further important to have in mind that \ion{Mg}{2}~h\&k emission exists all around the SOT threads, as evidenced by Fig.~1 in Paper~1. This, together with the existence of dynamical coherence, may suggest a scenario in which the threads belong to a larger inhomogeneous flux tube (which would exist over the distance in which coherence is detected). Such a scenario is discussed in section~\ref{spatial}.

\subsection{Phase mixing}

In our simulations two important mechanisms for small-scale generation are present: KHI and phase mixing. However, the timescales between KHI and phase mixing are different when it comes to produce small scales. Indeed, from one hand, over several periods phase mixing generates multiple very small-scale layers of counter-streaming flows located around the resonant layer. This also occurs in the case of an initially inhomogeneous loop \citep{Terradas_2008ApJ...679.1611T}. By itself, phase mixing would then result in a broadening of the line width, a decrease of the Doppler signal and a loss of coherence \citep[this is clearly seen after 6 full periods in][]{Terradas_2008ApJ...679.1611T}. Such behavior is also confirmed by our numerical model. Phase mixing produces the out-of-phase behavior rapidly (due to the change of the Alfv\'en speed across the boundary layer), while its generation of smaller scales (counter-flows) is produced on longer time scales. Accordingly, a decrease of the Doppler signal is clear when comparing the end stages with the first stages of the simulation. On the other hand, the small scales produced by KHI occur rapidly (within a period) and line broadening from KHI is readily produced, as seen in Fig.~\ref{fig_idw}. Importantly, the small-scale vortex flows produced by the KHI near the boundary receive momentum from the larger scale azimuthal flow from the resonance, whose characteristics therefore dominate the average signal picked up by the slits, thus avoiding the loss of Doppler signal.

\subsection{Effect of spatial resolution - KHI vortices as threads?}\label{spatial}

The transverse sizes of the observed thread-like structures with SOT vary between 200 and 400~km approximately. As discussed previously, in phase oscillation of threads, both in the POS motion and in the associated LOS velocity, exists (movies~$3-6$ and Fig.~5 in Paper~1), suggesting that some dynamic coherence exists in the transverse direction with a length scale of a few 1,000~km or so. Assuming a resolving power of $0.1~R$ for SOT and given that the Ca II H emission is similar to that of Mg II k, but optically thin with little opacity (as shown in Fig.~7 of Paper~1), we would expect an SOT time slice to be similar to Fig.~\ref{fig_ang_mg2}. As shown by the figure, in this case SOT would be resolving the internal structure of the prominence flux tubes. The fact that \ion{Mg}{2}~h\&k emission exists all around the SOT threads (as shown by Fig.~1 in Paper~1) further suggests that these structures may belong to a larger flux tube. It is therefore reasonable to consider that the observed thread-like structure would be the result of KHI vortices. Keeping roughly the ratio between the spatial resolutions of SOT and IRIS (a ratio of $2-3$), in this case the spatial resolution of IRIS would be $0.3~R$. Fig.~6 in Paper~1 would therefore apply to this case, meaning that the observed characteristic coherence in the transverse direction and the phase difference between the transverse displacement in the POS and the LOS velocity can be explained.

As shown in Figs.~\ref{fig_idw} and \ref{fig_ang_mg2} the KHI vortices are short lived (on the order of a period or so). This short lifetime can explain the observed fast disappearance of threads and makes the damping hard to observe in the numerical model as well. Notice that the damping of the transverse POS motion of the flux tube is only clearly present at the beginning of the simulation. This is because of the fast damping of the quasi-mode, which also holds for an inhomogeneous multi-threaded loop \citep{Terradas_2008ApJ...679.1611T,Pascoe_etal_2011ApJ...731...73P}. The presence of KHI further complicates the detection of the overall damping due to the large amount of thread-like structures and line-of-sight superposition that occurs (please keep in mind that, as suggested by our numerical model the \ion{Ca}{2}~H line may be significantly optically thin and therefore a significant superposition of such threads is expected in the SOT images). Furthermore, the KHI is fed by resonant absorption and the KHI vortices are partly embedded in the azimuthal (dipole-like) flow around the boundary layer that results from the resonance (movie~1). This implies that the oscillations of the KHI threads can increase in amplitude and that the damping is much weaker than that of the POS motion of the flux tube since the energy is continuously fed in via resonant absorption. This is especially so since at the low resolution of the current instruments (compared to the numerical resolution) and since they move coherently to a large extent due to the large scale azimuthal flow in which they are embedded
each thick thread is actually an unresolved ensemble of threads. This behavior can be seen in Figs.~\ref{fig_idw} and \ref{fig_ang_mg2} for \ion{Mg}{2}~k and in Fig.~\ref{fig_ang_si4} for the \ion{Si}{4} line, and matches well the observations (section~\ref{damping}). The damping of the ensemble of threads is only clearly observed when comparing the general ensemble of oscillations by the threads at the beginning and at the end of the simulation, and depends on the continuous generation of the KHI at each oscillation and the cascading of energy towards the lower scales. The former is limited by the energy provided by the resonance, and the latter is guaranteed by the high Lundquist number in the corona. Note that being able to resolve the KHI threads does not mean that we are fully resolving the boundary layer of the flux tube. The KHI threads retain in general the same phase with the POS motion of the flux tube, as shown in Figs.~\ref{fig_idw} and \ref{fig_ang_mg2}. The detected Doppler signal, coming from a coarser instrument than the imaging signal, does not correspond to the individual signals of each thread but rather to the average signal of the ensemble of threads, as shown in Fig.~6 of Paper~1.

Despite the many observational facts supporting the interpretation of KHI vortices as threads, we must also consider the case in which we are not resolving the oscillating flux tube. In this case, we set a spatial resolution of $0.5~R$ and $1.0~R$ for SOT and IRIS, respectively. As can be seen in the top left panel of Fig.~\ref{fig_coarse}, little internal structure is obtained, which suggests that the observed thread-like structure is not related to KHI in this case. However, the panels on the right hand side clearly show that the characteristic phase difference is maintained. Interestingly, even at this coarse resolution the ensemble of threads can be picked up in the hotter \ion{Si}{4} line as a single large thread (shown in the bottom left panel). This single `thick' thread produced by the ensemble of KHI threads conserves a very similar oscillation phase as the \ion{Mg}{2}~k intensity oscillation (and is therefore out of phase with the LOS \ion{Mg}{2}~k velocity), shows little damping due to the momentum transfer from the azimuthal waves, and thus provides good agreement with the observations. 

\section{Conclusions}\label{conclusions}

In this work we combined high resolution observations with Hinode and IRIS (Paper~1) with advanced 3D MHD numerical resolutions coupled with appropriate forward modeling to present the observational signatures (imaging and spectroscopic) of transverse MHD waves in prominence flux tubes. Resonant absorption rapidly produces characteristic out-of-phase behavior between the POS motions (produced by the kink quasi-mode) and the LOS velocity (produced by the azimuthal flows generated by the torsional Alfv\'en waves). The azimuthal flows from the resonance further produce localized line broadening at the boundaries of the flux tube (also through the process of phase mixing). These features would be invisible to current imagers and spectrometers if it were not for a secondary phenomenon that can accompany transverse MHD waves: KHI instabilities rapidly produced in timescales of one period by the velocity shear at the boundaries of the flux tube. A complex interplay between resonant absorption and the KHI occurs in which the KHI extracts the energy from the resonant layer and dumps it into heat through viscous and ohmic dissipation at the generated vortices and current sheets. The vortices, combined with LOS projection effects (especially for optically thin emission) lead to thread-like structure within the flux tube with enhanced emissivity in chromospheric lines, similar to the coronal case \citep{Antolin_2014ApJ...787L..22A}. The large vortices rapidly degenerate into turbulent flows that propagate inwardly and enlarge the transition boundary of the flux tube between the core and the external medium. A major result is that since the KHI vortices receive momentum from the resonance flows the out-of-phase behaviour and line broadening from the resonance are transferred to the enlarged transition boundary, therefore making these phenomena observationally detectable. We have shown that the observational signatures from this model match well the observations. 

In the transverse MHD wave interpretation that we have put forward for explaining the observations we have considered two different scenarios depending on whether the flux tube in which the wave exists is resolved or not. The first case considers the observed threads as independent flux tubes for which we are not able to resolve their internal structure. Each thread undergoes a kink oscillation and the POS motion and LOS velocities we observe are associated to each kink separately. A sketch of this case is shown in Fig.~8 of Paper~1, \textit{panel A}. In this work we have considered this case modeling a single flux tube and shown that the observed characteristic coherence over a significant transverse scale and the out-of-phase relation between the transverse displacement in the POS and the LOS velocity can be explained. The obtained timescales for the heating of the flux tube in our model match well the variability in the emission that can be associated to heating in the observations. We have shown that these features are numerically robust. Indeed, when increasing the resolution and Lundquist numbers an increase of small-scale vortices and current sheets is obtained, which enhances the heating due to KHI but keeps the large-scale dynamics unaltered. The results are also observationally robust, being to a large extent independent of the optical thickness of the material and the LOS angle (except in the completely optically thick case and very shallow angles). The observed dynamic coherence and out-of-phase relation are an imprint of resonant absorption and depend on the KHI only for widespread heating and thus detectability. At high spatial resolution and in the case of a well-defined flux tube the turbulence generated by the KHI is reflected in an increase of line broadening towards the boundary, especially when observing along the direction of oscillation. The detection of those dynamic features over a wide range of LOS angles and spatial resolutions is thus greatly facilitated by the presence of this instability.

The second case corresponds to a scenario in which the internal structure of the flux tube is resolved. In the previous section we have suggested that in this case the observed threads could be the KHI vortices, and shown that the expected signatures match well with the observed SOT threads. A confirmation or rejection of this scenario may only be achieved with higher instrumental spatial resolution. Another suggested possibility is that of an inhomogeneous flux tube in which regions of high density within the flux tube host the observed threads. This case is pictured in Fig.~8 of Paper~1, \textit{panel B}, and we can refer to it as the multiple kink wave case. Each high density region can have its own resonant layer, leading to azimuthal flows around each thread. Numerical simulations of this case but in a fully coronal setup with a multi-stranded loop with an irregular cross-section clearly indicate that the ensemble of strands (threads in our case) evolve coherently, with a dominant frequency everywhere within the structure in a similar way as for the initially homogeneous loop case \citep{Terradas_2008ApJ...679.1611T,Pascoe_etal_2011ApJ...731...73P}. In other words, a large azimuthal flow also develops around the combined strand (thread) system. These results strongly support the presence of a larger scale azimuthal flow around the threads with the same kind of out-of-phase behavior between the POS motion and the LOS velocity. Our results for the initially homogeneous flux tube can therefore also be applied to this initially inhomogeneous flux tube case.

Lastly, we have shown that the heating produced by the KHI affects a large area around the (smaller) prominence core, bringing the temperature of the plasma in this layer from chromospheric to transition region values, an effect also reported in the coronal case \citep{Antolin_2014ApJ...787L..22A}. This signature is in turn picked up at any LOS angle by TR lines such as \ion{Si}{4}, leading to a gradual fading of the threads out of the chromospheric pass bands (except for the prominence core) and appearance in the TR pass bands. This effect is accompanied by an apparent thinning of the flux tube, especially in optically thin emission in chromospheric lines such as \ion{Ca}{2}~H. 

All the mechanisms presented here for prominence plasmas extrapolate also to the coronal scenario  \citep[of which only the imaging aspects are shown in][]{Antolin_2014ApJ...787L..22A}, and may also extrapolate to spicules, in which multi-threaded structure, transverse swaying and torsional motions have recently been reported \citep{DePontieu_2014Sci...346D.315D,Rouppe_2015ApJ...799L...3R,Skogsrud_2014ApJ...795L..23S}. The combination of resonant absorption and KHI may therefore play an important role in the chromospheric and coronal morphology and heating. The assessment of this mechanism as such is the subject of future work. 

\acknowledgments
P.A. and T.J.O were supported by JSPS KAKENHI Grant Number 25220703 (PI: S. Tsuneta).T.J.O was supported by JSPS KAKENHI Grant Number 25800120 (PI: T.J.O.).  B.D.P. was supported by NASA under contract NNG09FA40C (IRIS), NNX11AN98G and NNM12AB40P. T.V.D. was supported by FWO Vlaanderen's Odysseus programme, GOA-2015-014 (KU Leuven) and the IAP P7/08 CHARM (Belspo). Numerical computations were carried out on Cray XC30 at the Center for Computational Astrophysics, NAOJ. This work was (partly) carried out on the Solar Data Analysis System operated by the Astronomy Data Center in cooperation with the Hinode Science Center of the National Astronomical Observatory of Japan.

\bibliographystyle{aa}
\bibliography{ms.bbl}  

\begin{thebibliography}{102}
\expandafter\ifx\csname natexlab\endcsname\relax\def\natexlab#1{#1}\fi

\bibitem[{{Alfv{\'e}n}(1947)}]{Alfven_1947MNRAS.107..211A}
{Alfv{\'e}n}, H. 1947, \mnras, 107, 211

\bibitem[{{Andries} {et~al.}(2005){Andries}, {Arregui}, \&
  {Goossens}}]{Andries_2005ApJ...624L..57A}
{Andries}, J., {Arregui}, I., \& {Goossens}, M. 2005, \apjl, 624, L57

\bibitem[{{Andries} \& {Goossens}(2001)}]{Andries_2001AA...368.1083A}
{Andries}, J. \& {Goossens}, M. 2001, \aap, 368, 1083

\bibitem[{{Anfinogentov} {et~al.}(2013){Anfinogentov}, {Nistic{\`o}}, \&
  {Nakariakov}}]{Anfinogentov_2013AA...560A.107A}
{Anfinogentov}, S., {Nistic{\`o}}, G., \& {Nakariakov}, V.~M. 2013, \aap, 560,
  A107

\bibitem[{{Antolin} \& {Rouppe van der
  Voort}(2012)}]{Antolin_Rouppe_2012ApJ...745..152A}
{Antolin}, P. \& {Rouppe van der Voort}, L. 2012, \apj, 745, 152

\bibitem[{{Antolin} \& {Shibata}(2010)}]{Antolin_2010ApJ...712..494A}
{Antolin}, P. \& {Shibata}, K. 2010, \apj, 712, 494

\bibitem[{{Antolin} \& {Van
  Doorsselaere}(2013)}]{Antolin_VanDoorsselaere_2013AA...555A..74A}
{Antolin}, P. \& {Van Doorsselaere}, T. 2013, \aap, 555, A74

\bibitem[{{Antolin} \&
  {Verwichte}(2011)}]{Antolin_Verwichte_2011ApJ...736..121A}
{Antolin}, P. \& {Verwichte}, E. 2011, \apj, 736, 121

\bibitem[{{Antolin} {et~al.}(2014){Antolin}, {Yokoyama}, \& {Van
  Doorsselaere}}]{Antolin_2014ApJ...787L..22A}
{Antolin}, P., {Yokoyama}, T., \& {Van Doorsselaere}, T. 2014, \apjl, 787, L22

\bibitem[{{Arregui} \& {Ballester}(2011)}]{Arregui_2011SSRv..158..169A}
{Arregui}, I. \& {Ballester}, J.~L. 2011, \ssr, 158, 169

\bibitem[{{Arregui} {et~al.}(2012){Arregui}, {Oliver}, \&
  {Ballester}}]{Arregui_2012LRSP....9....2A}
{Arregui}, I., {Oliver}, R., \& {Ballester}, J.~L. 2012, Living Reviews in
  Solar Physics, 9, 2

\bibitem[{{Arregui} {et~al.}(2011){Arregui}, {Soler}, {Ballester}, \&
  {Wright}}]{Arregui_2011AA...533A..60A}
{Arregui}, I., {Soler}, R., {Ballester}, J.~L., \& {Wright}, A.~N. 2011, \aap,
  533, A60

\bibitem[{{Arregui} {et~al.}(2007){Arregui}, {Terradas}, {Oliver}, \&
  {Ballester}}]{Arregui_2007AA...466.1145A}
{Arregui}, I., {Terradas}, J., {Oliver}, R., \& {Ballester}, J.~L. 2007, \aap,
  466, 1145

\bibitem[{{Arregui} {et~al.}(2008){Arregui}, {Terradas}, {Oliver}, \&
  {Ballester}}]{Arregui_2008ApJ...682L.141A}
{Arregui}, I., {Terradas}, J., {Oliver}, R., \& {Ballester}, J.~L. 2008, \apjl,
  682, L141

\bibitem[{{Aschwanden} {et~al.}(1999){Aschwanden}, {Fletcher}, {Schrijver}, \&
  {Alexander}}]{Aschwanden_1999ApJ...520..880A}
{Aschwanden}, M.~J., {Fletcher}, L., {Schrijver}, C.~J., \& {Alexander}, D.
  1999, \apj, 520, 880

\bibitem[{{Aschwanden} {et~al.}(2003){Aschwanden}, {Nightingale}, {Andries},
  {Goossens}, \& {Van
  Doorsselaere}}]{Aschwanden_Nightingale_2003ApJ...598.1375A}
{Aschwanden}, M.~J., {Nightingale}, R.~W., {Andries}, J., {Goossens}, M., \&
  {Van Doorsselaere}, T. 2003, \apj, 598, 1375

\bibitem[{{Brooks} {et~al.}(2013){Brooks}, {Warren}, {Ugarte-Urra}, \&
  {Winebarger}}]{Brooks_2013ApJ...772L..19B}
{Brooks}, D.~H., {Warren}, H.~P., {Ugarte-Urra}, I., \& {Winebarger}, A.~R.
  2013, \apjl, 772, L19

\bibitem[{{De Moortel} \&
  {Nakariakov}(2012)}]{DeMoortel_Nakariakov_2012RSPTA.370.3193D}
{De Moortel}, I. \& {Nakariakov}, V.~M. 2012, Royal Society of London
  Philosophical Transactions Series A, 370, 3193

\bibitem[{{De Pontieu} {et~al.}(2007{\natexlab{a}}){De Pontieu}, {McIntosh},
  {Hansteen}, {Carlsson}, {Schrijver}, {Tarbell}, {Title}, {Shine}, {Suematsu},
  {Tsuneta}, {Katsukawa}, {Ichimoto}, {Shimizu}, \&
  {Nagata}}]{DePontieu_etal_2007PASJ...59S.655D}
{De Pontieu}, B., {McIntosh}, S., {Hansteen}, V.~H., {et~al.}
  2007{\natexlab{a}}, \pasj, 59, 655

\bibitem[{{De Pontieu} {et~al.}(2007{\natexlab{b}}){De Pontieu}, {McIntosh},
  {Carlsson}, {Hansteen}, {Tarbell}, {Schrijver}, {Title}, {Shine}, {Tsuneta},
  {Katsukawa}, {Ichimoto}, {Suematsu}, {Shimizu}, \&
  {Nagata}}]{DePontieu_2007Sci...318.1574D}
{De Pontieu}, B., {McIntosh}, S.~W., {Carlsson}, M., {et~al.}
  2007{\natexlab{b}}, Science, 318, 1574

\bibitem[{{De Pontieu} {et~al.}(2014{\natexlab{a}}){De Pontieu}, {Rouppe van
  der Voort}, {McIntosh}, {Pereira}, {Carlsson}, {Hansteen}, {Skogsrud},
  {Lemen}, {Title}, {Boerner}, {Hurlburt}, {Tarbell}, {Wuelser}, {De Luca},
  {Golub}, {McKillop}, {Reeves}, {Saar}, {Testa}, {Tian}, {Kankelborg},
  {Jaeggli}, {Kleint}, \& {Martinez-Sykora}}]{DePontieu_2014Sci...346D.315D}
{De Pontieu}, B., {Rouppe van der Voort}, L., {McIntosh}, S.~W., {et~al.}
  2014{\natexlab{a}}, Science, 346, D315

\bibitem[{{De Pontieu} {et~al.}(2014{\natexlab{b}}){De Pontieu}, {Title},
  {Lemen}, {Kushner}, {Akin}, {Allard}, {Berger}, {Boerner}, {Cheung}, {Chou},
  {Drake}, {Duncan}, {Freeland}, {Heyman}, {Hoffman}, {Hurlburt}, {Lindgren},
  {Mathur}, {Rehse}, {Sabolish}, {Seguin}, {Schrijver}, {Tarbell},
  {W{\"u}lser}, {Wolfson}, {Yanari}, {Mudge}, {Nguyen-Phuc}, {Timmons}, {van
  Bezooijen}, {Weingrod}, {Brookner}, {Butcher}, {Dougherty}, {Eder},
  {Knagenhjelm}, {Larsen}, {Mansir}, {Phan}, {Boyle}, {Cheimets}, {DeLuca},
  {Golub}, {Gates}, {Hertz}, {McKillop}, {Park}, {Perry}, {Podgorski},
  {Reeves}, {Saar}, {Testa}, {Tian}, {Weber}, {Dunn}, {Eccles}, {Jaeggli},
  {Kankelborg}, {Mashburn}, {Pust}, {Springer}, {Carvalho}, {Kleint}, {Marmie},
  {Mazmanian}, {Pereira}, {Sawyer}, {Strong}, {Worden}, {Carlsson}, {Hansteen},
  {Leenaarts}, {Wiesmann}, {Aloise}, {Chu}, {Bush}, {Scherrer}, {Brekke},
  {Martinez-Sykora}, {Lites}, {McIntosh}, {Uitenbroek}, {Okamoto}, {Gummin},
  {Auker}, {Jerram}, {Pool}, \& {Waltham}}]{DePontieu_2014SoPh..289.2733D}
{De Pontieu}, B., {Title}, A.~M., {Lemen}, J.~R., {et~al.} 2014{\natexlab{b}},
  \solphys, 289, 2733

\bibitem[{{Doorsselaere} \&
  {Nakariakov}(2008)}]{VanDoorsselaere_2008ASPC..397...58D}
{Doorsselaere}, T.~V. \& {Nakariakov}, V.~M. 2008, in Astronomical Society of
  the Pacific Conference Series, Vol. 397, First Results From Hinode, ed. S.~A.
  {Matthews}, J.~M. {Davis}, \& L.~K. {Harra}, 58

\bibitem[{{Doschek} {et~al.}(2007){Doschek}, {Mariska}, {Warren}, {Brown},
  {Culhane}, {Hara}, {Watanabe}, {Young}, \&
  {Mason}}]{Doschek_etal_2007ApJ...667L.109D}
{Doschek}, G.~A., {Mariska}, J.~T., {Warren}, H.~P., {et~al.} 2007, \apjl, 667,
  L109

\bibitem[{{Erd{\'e}lyi} \& {Taroyan}(2008)}]{Erdelyi_2008AA...489L..49E}
{Erd{\'e}lyi}, R. \& {Taroyan}, Y. 2008, \aap, 489, L49

\bibitem[{{Fujimoto} {et~al.}(2006){Fujimoto}, {Nakamura}, \&
  {Hasegawa}}]{Fujimoto_2006SSRv..122....3F}
{Fujimoto}, M., {Nakamura}, T.~K.~M., \& {Hasegawa}, H. 2006, \ssr, 122, 3

\bibitem[{{Fujimoto} \& {Terasawa}(1994)}]{Fujimoto_1994JGR....99.8601F}
{Fujimoto}, M. \& {Terasawa}, T. 1994, \jgr, 99, 8601

\bibitem[{{Goossens} {et~al.}(2006){Goossens}, {Andries}, \&
  {Arregui}}]{Goossens_2006RSPTA.364..433G}
{Goossens}, M., {Andries}, J., \& {Arregui}, I. 2006, Royal Society of London
  Philosophical Transactions Series A, 364, 433

\bibitem[{{Goossens} {et~al.}(2002){Goossens}, {Andries}, \&
  {Aschwanden}}]{Goossens_2002AA...394L..39G}
{Goossens}, M., {Andries}, J., \& {Aschwanden}, M.~J. 2002, \aap, 394, L39

\bibitem[{{Goossens} {et~al.}(2012){Goossens}, {Andries}, {Soler}, {Van
  Doorsselaere}, {Arregui}, \& {Terradas}}]{Goossens_2012ApJ...753..111G}
{Goossens}, M., {Andries}, J., {Soler}, R., {et~al.} 2012, \apj, 753, 111

\bibitem[{{Goossens} {et~al.}(2011){Goossens}, {Erd{\'e}lyi}, \&
  {Ruderman}}]{Goossens_2011SSRv..158..289G}
{Goossens}, M., {Erd{\'e}lyi}, R., \& {Ruderman}, M.~S. 2011, \ssr, 158, 289

\bibitem[{{Goossens} {et~al.}(1992){Goossens}, {Hollweg}, \&
  {Sakurai}}]{Goossens_1992SoPh..138..233G}
{Goossens}, M., {Hollweg}, J.~V., \& {Sakurai}, T. 1992, \solphys, 138, 233

\bibitem[{{Goossens} {et~al.}(2014){Goossens}, {Soler}, {Terradas}, {Van
  Doorsselaere}, \& {Verth}}]{Goossens_2014ApJ...788....9G}
{Goossens}, M., {Soler}, R., {Terradas}, J., {Van Doorsselaere}, T., \&
  {Verth}, G. 2014, \apj, 788, 9

\bibitem[{{Gruszecki} {et~al.}(2008){Gruszecki}, {Murawski}, \&
  {Ofman}}]{Gruszecki_2008AA...488..757G}
{Gruszecki}, M., {Murawski}, K., \& {Ofman}, L. 2008, \aap, 488, 757

\bibitem[{{Heyvaerts} \& {Priest}(1983)}]{Heyvaerts_1983AA...117..220H}
{Heyvaerts}, J. \& {Priest}, E.~R. 1983, \aap, 117, 220

\bibitem[{{Hillier} {et~al.}(2013){Hillier}, {Morton}, \&
  {Erd{\'e}lyi}}]{Hillier_2013ApJ...779L..16H}
{Hillier}, A., {Morton}, R.~J., \& {Erd{\'e}lyi}, R. 2013, \apjl, 779, L16

\bibitem[{{Hollweg} {et~al.}(1990){Hollweg}, {Yang}, {Cadez}, \&
  {Gakovic}}]{Hollweg_1990ApJ...349..335H}
{Hollweg}, J.~V., {Yang}, G., {Cadez}, V.~M., \& {Gakovic}, B. 1990, \apj, 349,
  335

\bibitem[{{Ionson}(1978)}]{Ionson_1978ApJ...226..650I}
{Ionson}, J.~A. 1978, \apj, 226, 650

\bibitem[{{Kappraff} \&
  {Tataronis}(1977)}]{Kappraff_Tataronis_1977JPlPh..18..209K}
{Kappraff}, J.~M. \& {Tataronis}, J.~A. 1977, Journal of Plasma Physics, 18,
  209

\bibitem[{{Karpen} {et~al.}(1993){Karpen}, {Antiochos}, {Dahlburg}, \&
  {Spicer}}]{Karpen_1993ApJ...403..769K}
{Karpen}, J.~T., {Antiochos}, S.~K., {Dahlburg}, R.~B., \& {Spicer}, D.~S.
  1993, \apj, 403, 769

\bibitem[{{Kosugi} {et~al.}(2007){Kosugi}, {Matsuzaki}, {Sakao}, {Shimizu},
  {Sone}, {Tachikawa}, {Hashimoto}, {Minesugi}, {Ohnishi}, {Yamada}, {Tsuneta},
  {Hara}, {Ichimoto}, {Suematsu}, {Shimojo}, {Watanabe}, {Shimada}, {Davis},
  {Hill}, {Owens}, {Title}, {Culhane}, {Harra}, {Doschek}, \&
  {Golub}}]{Kosugi_2007SoPh..243....3K}
{Kosugi}, T., {Matsuzaki}, K., {Sakao}, T., {et~al.} 2007, \solphys, 243, 3

\bibitem[{{Kudoh} {et~al.}(1999){Kudoh}, {Matsumoto}, \&
  {Shibata}}]{Kudoh_1999_CFD.8}
{Kudoh}, T., {Matsumoto}, R., \& {Shibata}, K. 1999, Computational Fluid
  Dynamics Journal, 8, 56

\bibitem[{{Kuridze} {et~al.}(2012){Kuridze}, {Morton}, {Erd{\'e}lyi},
  {Dorrian}, {Mathioudakis}, {Jess}, \& {Keenan}}]{Kuridze_2012ApJ...750...51K}
{Kuridze}, D., {Morton}, R.~J., {Erd{\'e}lyi}, R., {et~al.} 2012, \apj, 750, 51

\bibitem[{{Lapenta} \& {Knoll}(2003)}]{Lapenta_2003SoPh..214..107L}
{Lapenta}, G. \& {Knoll}, D.~A. 2003, \solphys, 214, 107

\bibitem[{{Lin}(2011)}]{Lin_2011SSRv..158..237L}
{Lin}, Y. 2011, \ssr, 158, 237

\bibitem[{{Lin} {et~al.}(2009){Lin}, {Soler}, {Engvold}, {Ballester},
  {Langangen}, {Oliver}, \& {Rouppe van der Voort}}]{Lin_2009ApJ...704..870L}
{Lin}, Y., {Soler}, R., {Engvold}, O., {et~al.} 2009, \apj, 704, 870

\bibitem[{{Mackay} {et~al.}(2010){Mackay}, {Karpen}, {Ballester}, {Schmieder},
  \& {Aulanier}}]{Mackay_2010SSRv..151..333M}
{Mackay}, D.~H., {Karpen}, J.~T., {Ballester}, J.~L., {Schmieder}, B., \&
  {Aulanier}, G. 2010, \ssr, 151, 333

\bibitem[{{Mart{\'{\i}}nez-Sykora} {et~al.}(2012){Mart{\'{\i}}nez-Sykora}, {De
  Pontieu}, \& {Hansteen}}]{MartinezSykora_2012ApJ...753..161M}
{Mart{\'{\i}}nez-Sykora}, J., {De Pontieu}, B., \& {Hansteen}, V. 2012, \apj,
  753, 161

\bibitem[{{Mathioudakis} {et~al.}(2013){Mathioudakis}, {Jess}, \&
  {Erd{\'e}lyi}}]{Mathioudakis_2013SSRv..175....1M}
{Mathioudakis}, M., {Jess}, D.~B., \& {Erd{\'e}lyi}, R. 2013, \ssr, 175, 1

\bibitem[{{Matsumoto} \& {Suzuki}(2014)}]{Matsumoto_Suzuki_2014MNRAS.440..971M}
{Matsumoto}, T. \& {Suzuki}, T.~K. 2014, \mnras, 440, 971

\bibitem[{{McIntosh} {et~al.}(2011){McIntosh}, {de Pontieu}, {Carlsson},
  {Hansteen}, {Boerner}, \& {Goossens}}]{McIntosh_2011Natur.475..477M}
{McIntosh}, S.~W., {de Pontieu}, B., {Carlsson}, M., {et~al.} 2011, \nat, 475,
  477

\bibitem[{{Morton} {et~al.}(2012){Morton}, {Verth}, {Jess}, {Kuridze},
  {Ruderman}, {Mathioudakis}, \& {Erdelyi}}]{Morton_1012NatCommun...3...1315}
{Morton}, R.~J., {Verth}, G., {Jess}, D.~B., {et~al.} 2012, Nat Commun, 3, 1315

\bibitem[{{Nakariakov} \& {Ofman}(2001)}]{Nakariakov_Ofman_2001AA...372L..53N}
{Nakariakov}, V.~M. \& {Ofman}, L. 2001, \aap, 372, L53

\bibitem[{{Nakariakov} {et~al.}(1999){Nakariakov}, {Ofman}, {Deluca},
  {Roberts}, \& {Davila}}]{Nakariakov_1999Sci...285..862N}
{Nakariakov}, V.~M., {Ofman}, L., {Deluca}, E.~E., {Roberts}, B., \& {Davila},
  J.~M. 1999, Science, 285, 862

\bibitem[{{Ofman}(2009)}]{Ofman_2009ApJ...694..502O}
{Ofman}, L. 2009, \apj, 694, 502

\bibitem[{{Ofman} \& {Davila}(1995)}]{Ofman_Davila_1995JGR...10023427O}
{Ofman}, L. \& {Davila}, J.~M. 1995, \jgr, 100, 23427

\bibitem[{{Ofman} {et~al.}(1994){Ofman}, {Davila}, \&
  {Steinolfson}}]{Ofman_1994GeoRL..21.2259O}
{Ofman}, L., {Davila}, J.~M., \& {Steinolfson}, R.~S. 1994, \grl, 21, 2259

\bibitem[{{Ofman} {et~al.}(1998){Ofman}, {Kucera}, {Mouradian}, \&
  {Poland}}]{Ofman_1998SoPh..183...97O}
{Ofman}, L., {Kucera}, T.~A., {Mouradian}, Z., \& {Poland}, A.~I. 1998,
  \solphys, 183, 97

\bibitem[{{Ofman} \& {Wang}(2008)}]{Ofman_Wang_2008AA...482L...9O}
{Ofman}, L. \& {Wang}, T.~J. 2008, \aap, 482, L9

\bibitem[{{Okamoto} {et~al.}(2015){Okamoto}, {Antolin}, {De Pontieu},
  {Uitenbroek}, {Van Doorsselaere}, \& {Yokoyama}}]{Okamoto_etal_2015}
{Okamoto}, T.~J., {Antolin}, P., {De Pontieu}, B., {et~al.} 2015, Submitted to
  ApJ

\bibitem[{{Okamoto} {et~al.}(2007){Okamoto}, {Tsuneta}, {Berger}, {Ichimoto},
  {Katsukawa}, {Lites}, {Nagata}, {Shibata}, {Shimizu}, {Shine}, {Suematsu},
  {Tarbell}, \& {Title}}]{Okamoto_2007Sci...318.1577O}
{Okamoto}, T.~J., {Tsuneta}, S., {Berger}, T.~E., {et~al.} 2007, Science, 318,
  1577

\bibitem[{{Oliver}(2009)}]{Oliver_2009SSRv..149..175O}
{Oliver}, R. 2009, \ssr, 149, 175

\bibitem[{{Parnell} \& {De
  Moortel}(2012)}]{Parnell_DeMoortel_2012RSPTA.370.3217P}
{Parnell}, C.~E. \& {De Moortel}, I. 2012, Royal Society of London
  Philosophical Transactions Series A, 370, 3217

\bibitem[{{Pascoe} {et~al.}(2010){Pascoe}, {Wright}, \& {De
  Moortel}}]{Pascoe_2010ApJ...711..990P}
{Pascoe}, D.~J., {Wright}, A.~N., \& {De Moortel}, I. 2010, \apj, 711, 990

\bibitem[{{Pascoe} {et~al.}(2011){Pascoe}, {Wright}, \& {De
  Moortel}}]{Pascoe_etal_2011ApJ...731...73P}
{Pascoe}, D.~J., {Wright}, A.~N., \& {De Moortel}, I. 2011, \apj, 731, 73

\bibitem[{{Peter} {et~al.}(2013){Peter}, {Bingert}, {Klimchuk}, {de Forest},
  {Cirtain}, {Golub}, {Winebarger}, {Kobayashi}, \&
  {Korreck}}]{Peter_2013AA...556A.104P}
{Peter}, H., {Bingert}, S., {Klimchuk}, J.~A., {et~al.} 2013, \aap, 556, A104

\bibitem[{{Poedts} {et~al.}(1990){Poedts}, {Goossens}, \&
  {Kerner}}]{Poedts_1990CoPhC..59...95P}
{Poedts}, S., {Goossens}, M., \& {Kerner}, W. 1990, Computer Physics
  Communications, 59, 95

\bibitem[{{Poedts} \& {Kerner}(1991)}]{Poedts_Kerner_1991PhRvL..66.2871P}
{Poedts}, S. \& {Kerner}, W. 1991, Physical Review Letters, 66, 2871

\bibitem[{{Poedts} {et~al.}(1997){Poedts}, {Toth}, {Belien}, \&
  {Goedbloed}}]{Poedts_1997SoPh..172...45P}
{Poedts}, S., {Toth}, G., {Belien}, A.~J.~C., \& {Goedbloed}, J.~P. 1997,
  \solphys, 172, 45

\bibitem[{{Reale}(2010)}]{Reale_2010LRSP....7....5R}
{Reale}, F. 2010, Living Reviews in Solar Physics, 7, 5

\bibitem[{{Rouppe van der Voort} {et~al.}(2015){Rouppe van der Voort}, {De
  Pontieu}, {Pereira}, {Carlsson}, \& {Hansteen}}]{Rouppe_2015ApJ...799L...3R}
{Rouppe van der Voort}, L., {De Pontieu}, B., {Pereira}, T.~M.~D., {Carlsson},
  M., \& {Hansteen}, V. 2015, \apjl, 799, L3

\bibitem[{{Rybicki} \& {Hummer}(1992)}]{Rybicki_1992AA...262..209R}
{Rybicki}, G.~B. \& {Hummer}, D.~G. 1992, \aap, 262, 209

\bibitem[{{Sakurai} {et~al.}(1991){Sakurai}, {Goossens}, \&
  {Hollweg}}]{Sakurai_1991SoPh..133..227S}
{Sakurai}, T., {Goossens}, M., \& {Hollweg}, J.~V. 1991, \solphys, 133, 227

\bibitem[{{Schmieder} {et~al.}(2013){Schmieder}, {Kucera}, {Knizhnik}, {Luna},
  {Lopez-Ariste}, \& {Toot}}]{Schmieder_2013ApJ...777..108S}
{Schmieder}, B., {Kucera}, T.~A., {Knizhnik}, K., {et~al.} 2013, \apj, 777, 108

\bibitem[{{Scullion} {et~al.}(2014){Scullion}, {Rouppe van der Voort},
  {Wedemeyer}, \& {Antolin}}]{Scullion_2014ApJ...797...36S}
{Scullion}, E., {Rouppe van der Voort}, L., {Wedemeyer}, S., \& {Antolin}, P.
  2014, \apj, 797, 36

\bibitem[{{Skogsrud} {et~al.}(2014){Skogsrud}, {Rouppe van der Voort}, \& {De
  Pontieu}}]{Skogsrud_2014ApJ...795L..23S}
{Skogsrud}, H., {Rouppe van der Voort}, L., \& {De Pontieu}, B. 2014, \apjl,
  795, L23

\bibitem[{{Soler} {et~al.}(2012){Soler}, {Ruderman}, \&
  {Goossens}}]{Soler_2012AA...546A..82S}
{Soler}, R., {Ruderman}, M.~S., \& {Goossens}, M. 2012, \aap, 546, A82

\bibitem[{{Soler} {et~al.}(2010){Soler}, {Terradas}, {Oliver}, {Ballester}, \&
  {Goossens}}]{Soler_2010ApJ...712..875S}
{Soler}, R., {Terradas}, J., {Oliver}, R., {Ballester}, J.~L., \& {Goossens},
  M. 2010, \apj, 712, 875

\bibitem[{{Southwood} \& {Hughes}(1983)}]{Southwood_Hughes_1983SSRv...35..301S}
{Southwood}, D.~J. \& {Hughes}, W.~J. 1983, \ssr, 35, 301

\bibitem[{{Tandberg-Hanssen}(1995)}]{Tandberg-Hansen_1995ASSL..199.....T}
{Tandberg-Hanssen}, E., ed. 1995, Astrophysics and Space Science Library, Vol.
  199, {The nature of solar prominences}

\bibitem[{{Terradas} {et~al.}(2008{\natexlab{a}}){Terradas}, {Andries},
  {Goossens}, {Arregui}, {Oliver}, \&
  {Ballester}}]{Terradas_2008ApJ...687L.115T}
{Terradas}, J., {Andries}, J., {Goossens}, M., {et~al.} 2008{\natexlab{a}},
  \apjl, 687, L115

\bibitem[{{Terradas} {et~al.}(2008{\natexlab{b}}){Terradas}, {Arregui},
  {Oliver}, \& {Ballester}}]{Terradas_2008ApJ...678L.153T}
{Terradas}, J., {Arregui}, I., {Oliver}, R., \& {Ballester}, J.~L.
  2008{\natexlab{b}}, \apjl, 678, L153

\bibitem[{{Terradas} {et~al.}(2008{\natexlab{c}}){Terradas}, {Arregui},
  {Oliver}, {Ballester}, {Andries}, \&
  {Goossens}}]{Terradas_2008ApJ...679.1611T}
{Terradas}, J., {Arregui}, I., {Oliver}, R., {et~al.} 2008{\natexlab{c}}, \apj,
  679, 1611

\bibitem[{{Terradas} {et~al.}(2010){Terradas}, {Goossens}, \&
  {Ballai}}]{Terradas_2010AA...515A..46T}
{Terradas}, J., {Goossens}, M., \& {Ballai}, I. 2010, \aap, 515, A46

\bibitem[{{Terradas} {et~al.}(2006){Terradas}, {Oliver}, \&
  {Ballester}}]{Terradas_etal_2006ApJ...642..533T}
{Terradas}, J., {Oliver}, R., \& {Ballester}, J.~L. 2006, \apj, 642, 533

\bibitem[{{Threlfall} {et~al.}(2013){Threlfall}, {De Moortel}, {McIntosh}, \&
  {Bethge}}]{Threlfall_2013AA...556A.124T}
{Threlfall}, J., {De Moortel}, I., {McIntosh}, S.~W., \& {Bethge}, C. 2013,
  \aap, 556, A124

\bibitem[{{Thurgood} {et~al.}(2014){Thurgood}, {Morton}, \&
  {McLaughlin}}]{Thurgood_2014ApJ...790L...2T}
{Thurgood}, J.~O., {Morton}, R.~J., \& {McLaughlin}, J.~A. 2014, \apjl, 790, L2

\bibitem[{{Tian} {et~al.}(2012){Tian}, {McIntosh}, {Wang}, {Ofman}, {De
  Pontieu}, {Innes}, \& {Peter}}]{Tian_2012ApJ...759..144T}
{Tian}, H., {McIntosh}, S.~W., {Wang}, T., {et~al.} 2012, \apj, 759, 144

\bibitem[{{Tirry} \& {Goossens}(1996)}]{Tirry_Goossens_1996ApJ...471..501T}
{Tirry}, W.~J. \& {Goossens}, M. 1996, \apj, 471, 501

\bibitem[{{Tomczyk} \& {McIntosh}(2009)}]{Tomczyk_2009ApJ...697.1384T}
{Tomczyk}, S. \& {McIntosh}, S.~W. 2009, \apj, 697, 1384

\bibitem[{{Tomczyk} {et~al.}(2007){Tomczyk}, {McIntosh}, {Keil}, {Judge},
  {Schad}, {Seeley}, \& {Edmondson}}]{Tomczyk_2007Sci...317.1192T}
{Tomczyk}, S., {McIntosh}, S.~W., {Keil}, S.~L., {et~al.} 2007, Science, 317,
  1192

\bibitem[{{Tsuneta} {et~al.}(2008){Tsuneta}, {Ichimoto}, {Katsukawa}, {Nagata},
  {Otsubo}, {Shimizu}, {Suematsu}, {Nakagiri}, {Noguchi}, {Tarbell}, {Title},
  {Shine}, {Rosenberg}, {Hoffmann}, {Jurcevich}, {Kushner}, {Levay}, {Lites},
  {Elmore}, {Matsushita}, {Kawaguchi}, {Saito}, {Mikami}, {Hill}, \&
  {Owens}}]{Tsuneta_2008SoPh..249..167T}
{Tsuneta}, S., {Ichimoto}, K., {Katsukawa}, Y., {et~al.} 2008, \solphys, 249,
  167

\bibitem[{{Uchida} \& {Kaburaki}(1974)}]{Uchida_1974SoPh...35..451U}
{Uchida}, Y. \& {Kaburaki}, O. 1974, \solphys, 35, 451

\bibitem[{{Uchimoto} {et~al.}(1991){Uchimoto}, {Strauss}, \&
  {Lawson}}]{Uchimoto_1991SoPh..134..111U}
{Uchimoto}, E., {Strauss}, H.~R., \& {Lawson}, W.~S. 1991, \solphys, 134, 111

\bibitem[{{Uitenbroek}(2001)}]{Uitenbroek_2001ApJ...557..389U}
{Uitenbroek}, H. 2001, \apj, 557, 389

\bibitem[{{van Ballegooijen} {et~al.}(2011){van Ballegooijen}, {Asgari-Targhi},
  {Cranmer}, \& {DeLuca}}]{vanBallegooijen_etal_2011ApJ...736....3V}
{van Ballegooijen}, A.~A., {Asgari-Targhi}, M., {Cranmer}, S.~R., \& {DeLuca},
  E.~E. 2011, \apj, 736, 3

\bibitem[{{Van Doorsselaere} {et~al.}(2004){Van Doorsselaere}, {Andries},
  {Poedts}, \& {Goossens}}]{VanDoorsselaere_2004ApJ...606.1223V}
{Van Doorsselaere}, T., {Andries}, J., {Poedts}, S., \& {Goossens}, M. 2004,
  \apj, 606, 1223

\bibitem[{{Van Doorsselaere} {et~al.}(2008){Van Doorsselaere}, {Nakariakov},
  {Young}, \& {Verwichte}}]{VanDoorsselaere_etal_2008AA...487L..17V}
{Van Doorsselaere}, T., {Nakariakov}, V.~M., {Young}, P.~R., \& {Verwichte}, E.
  2008, \aap, 487, L17

\bibitem[{{Verth} {et~al.}(2010){Verth}, {Terradas}, \&
  {Goossens}}]{Verth_2010ApJ...718L.102V}
{Verth}, G., {Terradas}, J., \& {Goossens}, M. 2010, \apjl, 718, L102

\bibitem[{{Vial} \& {Engvold}(2015)}]{Vial_Engvold_2015ASSL..415.....V}
{Vial}, J.-C. \& {Engvold}, O., eds. 2015, Astrophysics and Space Science
  Library, Vol. 415, {Solar Prominences}

\bibitem[{{Zhou} {et~al.}(2014){Zhou}, {Chen}, {Zhang}, \&
  {Fang}}]{Zhou_2014RAA....14..581Z}
{Zhou}, Y.-H., {Chen}, P.-F., {Zhang}, Q.-M., \& {Fang}, C. 2014, Research in
  Astronomy and Astrophysics, 14, 581

\bibitem[{{Ziegler} \& {Ulmschneider}(1997)}]{Ziegler_1997AA...327..854Z}
{Ziegler}, U. \& {Ulmschneider}, P. 1997, \aap, 327, 854

\end{thebibliography}

\end{document}